\documentclass[sigconf, natbib=true]{acmart}

\AtBeginDocument{%
  \providecommand\BibTeX{{%
    \normalfont B\kern-0.5em{\scshape i\kern-0.25em b}\kern-0.8em\TeX}}}

\setcopyright{acmcopyright}
\copyrightyear{2018}
\acmYear{2018}
\acmDOI{XXXXXXX.XXXXXXX}

\acmConference[Conference acronym 'XX]{Make sure to enter the correct
  conference title from your rights confirmation emai}{June 03--05,
  2018}{Woodstock, NY}
\acmPrice{15.00}
\acmISBN{978-1-4503-XXXX-X/18/06}

\usepackage{multirow}
\usepackage{footmisc}
\usepackage{subcaption,siunitx,booktabs}
\usepackage{caption}
\usepackage{bbm}
\usepackage{algorithmic}
\usepackage{autobreak}
\usepackage{amsmath}
\usepackage{witharrows}
\usepackage{enumitem}
\usepackage{siunitx}
\newcommand{\etal}{\textit{et al}.}
\newcommand{\ie}{\textit{i}.\textit{e}.}
\newcommand{\eg}{\textit{e}.\textit{g}.}
\newcommand{\wrt}{\textit{w}.\textit{r}.\textit{t}.} 
\newcommand{\wo}{\textit{w}.\textit{o}.}
\newcommand{\vpara}[1]{\vspace{0.05in}\noindent\textbf{#1 }}
\begin{document}

\title{Denoising Multi-modal Sequential Recommenders with Contrastive Learning}

\author{Dong Yao}

\affiliation{%
  \institution{Zhejiang Unversity}
  \country{Hangzhou,China}
}
\email{yaodongai@zju.edu.cn}

\author{Shengyu Zhang}
\affiliation{
  \institution{Zhejiang Unversity}
  \country{Hangzhou,China}
}
\email{sy_zhang@zju.edu.cn}

\author{Zhou Zhao}
\affiliation{
  \institution{Zhejiang Unversity}
  \country{Hangzhou,China}
}
\email{zhaozhou@zju.edu.cn}

\author{Jieming Zhu}
\affiliation{
  \institution{Huawei}
  \country{Shenzhen,China}
}
\email{jiemingzhu@ieee.org}

\author{Wenqiao Zhang}
\affiliation{
  \institution{Zhejiang Unversity}
  \country{Hangzhou,China}
}
\email{wenqiaozhang@zju.edu.cn}

\author{Rui Zhang}
\affiliation{
  \institution{ruizhang.info}
  \country{Shenzhen,China}
}
\email{rayteam@yeah.net}

\author{Xiaofei He}
\affiliation{
  \institution{Zhejiang Unversity}
  \country{Hangzhou,China}
}
\email{xiaofei_h@qq.com}

\author{Fei Wu}
\affiliation{
  \institution{Zhejiang Unversity}
  \country{Hangzhou,China}
}
\email{wufei@zju.edu.cn}

\renewcommand{\shortauthors}{Trovato and Tobin, et al.}

\begin{abstract}
There is a rapidly-growing research interest in engaging users with multi-modal data for accurate user modeling on recommender systems. Existing multimedia recommenders have achieved substantial improvements by incorporating various modalities and devising delicate modules. However, when users decide to interact with items, most of them do not fully read the content of all modalities. We refer to modalities that directly cause users' behaviors as point-of-interests, which are important aspects to capture users' interests. In contrast, modalities that do not cause users' behaviors are potential noises and might mislead the learning of a recommendation model. Not surprisingly, little research in the literature has been devoted to denoising such potential noises due to the inaccessibility of users' explicit feedback on their point-of-interests. 
To bridge the gap, we propose a weakly-supervised framework based on contrastive learning for denoising multi-modal recommenders (dubbed Demure). In a weakly-supervised manner, Demure circumvents the requirement of users' explicit feedback and identifies the noises by analyzing the modalities of all interacted items from a given user. 
Specifically, Demure identifies potential noises and point-of-interests by assessing the contributions of each modality on the final recommendation prediction, inspired by explainable machine learning techniques. Upon identification, Demure augments original user behavior sequences by strengthening point-of-interests and reducing potential noises to obtain augmented positive and negative multi-modal behavior sequences, respectively. With these self-augmented sequences, contrastive learning objectives are applied to learn more accurate and robust user representations. Through extensive experiments, we show that our Demure framework not only improves the performance of recommendation significantly but also provides an explanation of what are the key modalities of users' interests.
\end{abstract}




\maketitle

\section{Introduction}

\begin{figure}[!t] \begin{center}
    \includegraphics[width=0.6\columnwidth]{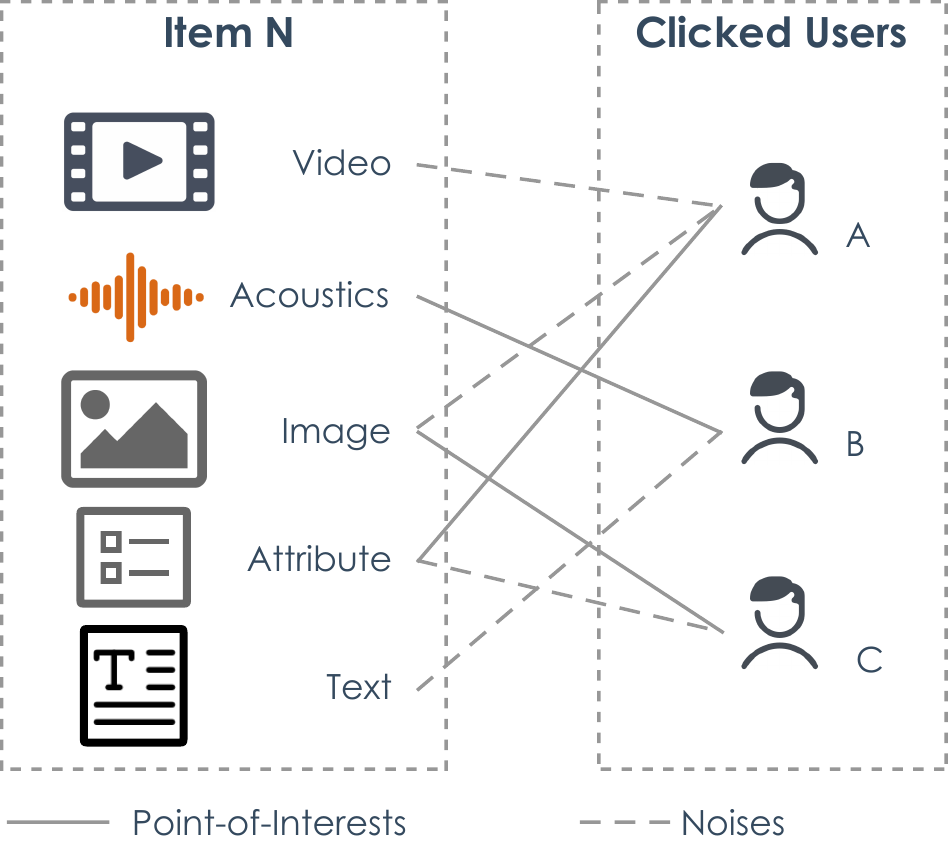}
    \caption{
    The click behaviors of different users might be driven by different modalities. Modalities that do not directly cause user' behaviors might mislead user representation learning.
    	}
\label{fig:firstpage}
\end{center} \end{figure}


For many online content-sharing websites, personalized recommendation serves as a vital service to ease the content discovery of users and improve their experience. Most items on content-sharing websites associate multi-modal features. For example, micro-videos provide visual frames, audio tracks, while textual descriptions \cite{Jiang_Wang_Wei_Gao_Wang_Nie_2020,Li_Liu_Yin_Cui_Xu_Nie_2019,Wang_Jiang_Xu_Xie_2019,Wei_Wang_Nie_He_Hong_Chua_2019}, and news articles come with images, videos, and textual titles \cite{Wang2020, Tian2021, Qi2021}. Analogously to the gains observed in other machine learning tasks, recent advances in content recommendation \cite{Fu2019,Huang_Fang_Qian_Sang_Li_Xu_2019,Wang_Jiang_Xu_Xie_2019,Wei_Wang_Nie_He_Chua_2020,ZhangJD0YCTHWHC21} have shown that engaging user modeling with multi-modal data can significantly improve recommendation performance.

Despite the fruitful progress made by previous studies, most of them solely deal with a single modality. Incorporating multi-modal information in recommendations is still a developing research area due to the following challenges: 1) how to bridge the semantic gap between different modalities? 2) how to devise user preference-oriented multi-modal modeling modules? Recently, \cite{Chen_Zhang_He_Nie_Liu_Chua_2017} employs a multi-level attention mechanism to aggregate multi-modal features for an item and multiple items within the behavior sequences. \cite{Wei_Wang_Nie_He_Hong_Chua_2019} constructs modal-specific user-item bipartite graphs and captures user preferences in each modality individually. Despite their significant advances, most existing studies neglect the noises in the multi-modal recommendation. Specifically, we do not mean that multi-modal features themselves are noisy or damaged. However, most users do not read the full content of all modalities when they decide to click or buy items. Actually, in most cases, users are attracted, and some particular modalities drive their behaviors. These modalities that directly cause users' behaviors are referred to as \textbf{point-of-interests} in this paper. 
For example, in Figure \ref{fig:firstpage}, user \texttt{A} may solely want to find a movie suitable for \texttt{kids and family} (\ie, the category attributes) but does not care about the content (\eg, video or image) when clicking on a movie. In such a case, modalities that do not directly cause the user's clicking behavior are potential noises for accurately modeling the preference of user \texttt{A}. 
Moreover, given the implicit feedback such as clicks and the ubiquitous distractions (\eg, discount information), the historically interacted items should also contain noisy ones that may be less representative of users' underlying preferences. \cite{Wang2021} demonstrates that the performance of clean training (without noisy samples) is much better than normal training (includes noises), which highlights the necessity to explore denoising in the recommendation. While we might collect some explicit feedback, such as like/follow signals in the item-level, we have significantly fewer signals in the modality-level indicating the point-of-interests that directly cause their behaviors.

In this paper, we propose a unified weakly-supervised framework for \textit{\textbf{de}}noising \textit{\textbf{mu}}lti-modal \textit{\textbf{re}}commenders in both item-level and modality-level, abbreviated as \textbf{Demure}. Demure is weakly supervised in the sense that it does not require any explicit feedback as supervision. Rather, it identifies point-of-interests and noises in items and modalities by assessing their contributions in predicting the next-item a user will interact with, which is a typical sequential recommendation formulation. Inspired by explainable machine learning techniques\cite{Selvaraju_Cogswell_Das_Vedantam_Parikh_Batra_2020}, we measure the contributions as the gradients estimated in the next-item prediction task. 
We intuitively view the modalities/items with higher contributions as the point-of-interests when predicting the next item during the training period.
Given the localized point-of-interests and potential noises, we apply contrastive learning to eliminate the effects of noises on final user embeddings. Specifically, we augment the original multi-modal behavior sequences by strengthening point-of-interests or reducing noises in both item level and modality level. Given that items/modalities that are point-of-interests are modified, multi-modal user sequences and the user encoder should jointly yield user embeddings that serve as negative views of the original user embeddings. Such negative user embeddings should primarily contain features with the noises. Pushing such negatives away from the original embeddings could help eliminate the effects of noises on the final user embeddings. Similarly, we could obtain positive user embeddings from the multi-modal behavior sequences with potential noises modified. As such, \textit{contrastive learning objectives} are designed to achieve accurate user preference modeling with multi-modal features based on these positives and negatives.

We conduct extensive experiments on several publicly available multi-modal recommendation benchmarks. We show in the experiments that Demure achieves the state-of-the-art results on multi-modal recommendation and provides explanations of user preferences on the modality- and item- levels. It is worth noting that the public news recommendation dataset MIND \cite{Wu_Qiao_Chen_Wu_Qi_Lian_Liu_Xie_Gao_Wu_et_2020} initially contains only one modality; thus, we crawl visual images to obtain a multi-modal version \footnote{We will release the dataset to promote further research and make our results reproducible}. Overall, the contributions of this work are threefold: 

\begin{itemize}[leftmargin=*]
    \item We identify the multi-modal noise issue in user representation learning and further make an initial effort to investigate denoising in recommendation without explicit signals indicating which modalities directly cause users' behaviors.
	\item We propose a weakly-supervised framework, named Demure, which identifies point-of-interests and potential noises by measuring modalities' and items' contributions in predicting the next item a user will interact with. Contrastive learning objectives are devised to learn robust user representations free of noises in both modality-level and item-level.
	\item We perform extensive experiments on the publicly available multi-modal recommendation benchmarks and achieve the state-of-the-art performance. We also contribute a multi-modal dataset by crawling the cover image of each news based on an open news recommendation dataset MIND.\footnote{All codes and datasets will be made publicly available to facilitate further improvements.}.
\end{itemize}

\section{Related Works}

\subsection{Multi-modal Recommendation}

Content-based recommendation's significance and challenging nature lend itself to great research interests and diverse investigations \cite{Chelliah_Biswas_Dhakad_2019,Du_Wang_He_Li_Tang_Chua_2020,Jiang_Wang_Liu_Nie_Duan_Xu_2019,Jiang_Wang_Wei_Gao_Wang_Nie_2020,Meng_Feng_He_Gao_Chua_2020,Verma_Gulati_Goel_Shah_2020,Yang_Xie_Wang_Yuan_Ding_Yan_2020,Yu_Gan_Wei_Cheng_Nie_2020,Zhang_Yuan_Li_Zhang_2019}. Compared with behavior-only preference modeling, a content-based recommendation has the following advantages. Firstly, contents convey multi-aspects and fine-grained details about items, which can enhance the representation of items and permit in-depth preference modeling of users. Secondly, content features can help mitigate the popularity bias on fresh items and are friendly for cold-start settings \cite{Chen_Zhang_He_Nie_Liu_Chua_2017}.

However, many works deal with a single modality, such as news title in news recommendation \cite{Wu_Wu_An_Huang_Huang_Xie_2019,Wu_Qiao_Chen_Wu_Qi_Lian_Liu_Xie_Gao_Wu_et_2020}and video feature in micro-video recommendations \cite{Jiang_Wang_Wei_Gao_Wang_Nie_2020,Li_Liu_Yin_Cui_Xu_Nie_2019}. How to encapsulate multiple modalities in recommendation remains largely unexplored in the research community. Recently, ACF \cite{Chen_Zhang_He_Nie_Liu_Chua_2017} progressively aggregates multiple modalities within an item and multiple items within a user sequence using attention mechanisms. MMGCN \cite{Wei_Wang_Nie_He_Hong_Chua_2019} constructs user-item bipartite graphs per modality and conducts modality-specific information propagations. Based on MMGCN, GRCN \cite{Wei_Wang_Nie_He_Chua_2020} improves it by further refining the modality-specific graphs by end-to-end learning. Unlike these works, we explicitly learn how to localize the point-of-interests of users and propose to denoise multi-modal learning with contrastive learning frameworks.

\subsection{Explainable Machine Learning}

The explanation for machine learning has increasingly attracted the attention of researchers, especially in the computer vision area. \cite{Zhang_Wu_Zhu_2018} propose an interpretable CNN in which each filter in a high conv-layer represents a specific object part that is different from conventional offline visualization. Grad-CAM \cite{Selvaraju_Cogswell_Das_Vedantam_Parikh_Batra_2020} utilizes the gradients to generate the class discriminative localization map that can highlight the critical regions in the image for the predicting class.

\subsection{Contrastive Learning}

The recent years have witnessed the great success of contrastive learning in a broad range of research domains \cite{He_Fan_Wu_Xie_Girshick_2020,Chen_Kornblith_Norouzi_Hinton_2020,Kang_Park_2020,Laskin_Srinivas_Abbeel_2020}. Such significant progress has also raised interest in the recommendation research field. Recently, Zhou \etal, \cite{zhou2020contrastive} deploy the contrastive learning objective into real-world recommender systems. They surprisingly find that such an objective can help alleviate popularity bias and theoretically prove that contrastive learning objective is closely related to more advanced debiased objectives such as inverse propensity score \cite{rosenbaum1983central}. Xie \etal, \cite{Xie_Sun_Liu_Gao_Ding_Cui_2020} propose the self-supervised learning framework for pretraining the user encoders from recommendation. Demure differs from these works in the goal, \ie, denoising multi-modal recommendation, and in the technique, \ie, contrastive learning based on the localization of point-of-interests in the modality-level and the item-level.

%

\section{Methods}

\subsection{Notations and Problems} \label{sec:problem}

A recommendation dataset generally contains two sets of entities, \ie, the user set $\mathcal{U}=\{u_1, u_2, \dots, u_{|\mathcal{U}|}\}$, and the item set $\mathcal{I}=\{i_1, i_2, \dots, i_{|\mathcal{I}|}\}$. $|\mathcal{U}|$ and $|\mathcal{I}|$ denote the number of users and items, respectively. Let $t$ represent the timestamp of interaction $<u,i,t>$ that happens, the interaction sequence of user $u$ can be formulated as $\mathcal{S} = \{ i_{u,t_1}, i_{u,t_2}, \dots, i_{u,t_{|\mathcal{S}|}} \}$, where $i_{u,t_1}$ denotes the item that user $u$ interacts with at timestamp $t_1$, and $|\mathcal{S}|$ denotes the length of behavior sequence. $t_1, t_2, \dots, t_{|\mathcal{S}|}$ are typically sorted in a chronological order. In a multi-modal setting, each item $i$ associates with a number of modalities $\{i^m\}_{m=1,\dots,M_i}$. The $M_i$ denotes the numbers of modalities of the item $i$. We use the public datasets where items have complete modalities, so the $M_i$ of all items belong to the same dataset are equal. For convenience, we simplify it as $M$. In detail, we apply a common setting of training sequential recommendation models \cite{Wang2021_Count, Zhang2021, Ren2020, Wang2020_Chorus, Wang2020_Next}.


\begin{figure*}[t] \begin{center}
    \includegraphics[width=\textwidth]{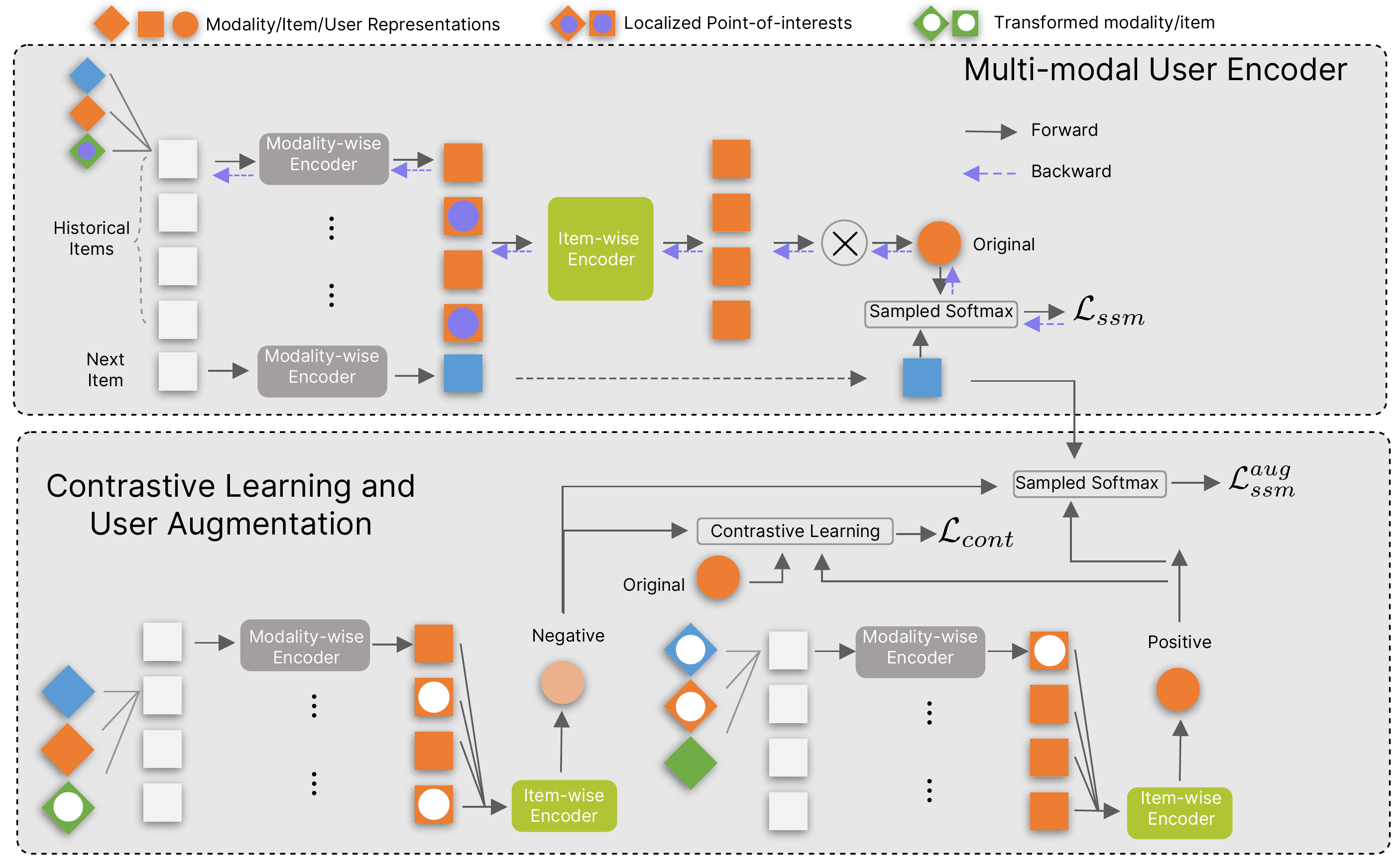}
    \caption{
    	A schematic of the proposed Demure framework. 
	}
\label{fig:schema}
\end{center} \end{figure*} 

\subsection{Multi-modal User Encoder} \label{sec:userencoder}

We firstly obtain the vectorial representations $\{\mathbf{i}^m\}_{m=1,\dots,|M|}$ of different modalities $\{i^m\}_{m=1,\dots,|M|}$ for each item $i$ using pre-trained modality-specific feature extractors\footnote{The details of feature extractors can be found in Section \ref{sec:expset}}. Given a user sequence $\mathcal{S} = \{ i_{t_1}, i_{t_2}, \dots, i_{t_{|\mathcal{S}|}} \}$, a user encoder aims at extracting preference patterns from the multi-modal features within historical behaviors and accordingly yields the user embeddings. The attention mechanism is a mature tool to model correlation and interaction among different modalities \cite{Chen2017, Lu2019} and items \cite{Chen2017}. We propose to leverage such benefit in the backbone of Demure. As illustrated in Figure 2, We choose to decompose the user encoder into two parts:

\vpara{Modality-wise Encoder} To bridge the modality gap and capture heterogeneous dependencies among different modalities, we leverage self-attention mechanism \cite{Vaswani_Shazeer_Parmar_Uszkoreit_Jones_Gomez_Kaiser_Polosukhin_2017} due to its effectiveness. Mathematically, given the features $\{\mathbf{i}^m\}_{m=1,\dots,|M|}$ of different modalities, we obtain the enhanced representation of $m$th modality as the following:
\begin{align}
	\beta_{m, n} &= \frac{\exp \left(s_{m,n}\right)}{\sum_{k=1}^{M} \exp \left(s_{m,k}\right)}, \mathrm{ where} \ \  s_{m,n}=f_m\left(\mathbf{i}^m\right)^{T} g_m\left(\mathbf{i}^n\right) \label{eq:selfatt_modality} \\
	\mathbf{\tilde i}^{m} &= \sum_{k=1}^{M} \beta_{m, k} h_m\left(\mathbf{i}^k\right)     
\end{align}
where $f_m$, $g_m$, and $h_m$ indicate linear transformations parameterized by weight matrices $\mathbf{W}_f$, $\mathbf{W}_g$, and $\mathbf{W}_h$, respectively. $s_{m,n}$ indicates the extent to which the model attends to the $n$th modality when representing the $m$th modality. $\mathbf{\tilde i}^{m}$ indicates the enhanced representation of the $m$th modality. To obtain the aggregated item embedding, we reuse the self-attention weights computed in Equation \ref{eq:selfatt_modality}. Specifically, the attention weight for modality $m$ can be computed as the following:
\begin{align}
	\beta_{m} = \frac{\sum_{n=1}^{M} \exp \left(s_{m,n}\right)}{ \sum_{m=1}^{M}\sum_{n=1}^{M} \exp \left(s_{m,n}\right) } \label{eq:attn_agg_modality}
\end{align}
\begin{sloppypar}
\noindent where $\sum_{n=1}^{M} \exp \left(s_{m,n}\right)$ indicate the extent to which the model attends to the $m$th modality when representing all modalities. Since items are represented by all modalities, it is reasonable to view such accumulated attention for modality $m$ normalized by $\sum_{m=1}^{M}\sum_{n=1}^{M} \exp \left(s_{m,n}\right)$ as the attention weight for representing the item. This design is also light-weight without no further learnable parameters. Therefore the aggregated item embedding can be $\mathbf{i}_t = \sum_{m=1}^{M} \beta_{m} \mathbf{\tilde i}^m$.
\end{sloppypar}

\vpara{Item-wise Encoder.}
Similarly, to capture heterogeneous dependencies among different items, we propose applying the self-attention mechanism in representing users. Specifically, given the vectorial representations of an user sequence $\mathbf{S} = \{ \mathbf{i}_{t_1}, \mathbf{i}_{t_2}, \dots, \mathbf{i}_{t_{|\mathcal{S}|}} \}$ where the representation of each item $\mathbf{i}_t$ is obtained from the modality-wise encoder. Instead of simply averaging them, we obtain the enhanced representation of all items within a sequence and reuse the attention weights to aggregate them into $\mathbf{i}$ in the same way as the modality-wise encoder. Afterward, we obtain the final representation $\mathbf{u}$ of user $u$ encoded by a feed-forward neural network (FFN):
\begin{align}
	\mathbf{u} &= \operatorname{FFN}\left(  \mathbf{i} \right)
\end{align}


\subsection{Prediction and Optimization} \label{sec:oriobj}

Unlike ranking models that encapsulate complex and learnable prediction layers and deal with a limited number of user-item pairs, matching models should retrieve top-K candidates from a massive item gallery and typically employ the non-parametric and efficient technique. In this paper, we use dot-product as the scoring function and use sample softmax for optimization, which is a widely used training objective and can be formulated as the following:
\begin{align}
 & \arg\min_\theta \mathcal{L}_{ssm} = \arg\min_\theta \frac{1}{|\mathcal{D}|}\sum_{(u,i^+)\in \mathcal{D}} -\log p_\theta'(i^+ \mid u),
    \quad \label{eq:obj} \\
 & p_\theta'(i^+\mid u)=\frac{\exp \left( \psi \left( u,i^+ \right) \right)}{\exp \left( \psi \left( u,i^+ \right) \right)+\sum_{i^- \in \mathcal{I}^-} \exp \left( \psi \left( x,i^- \right) \right)}
\end{align}
where $\mathcal{I}^-$ is the item set of sampled negatives. $i^+$ represents the target item. $\psi(u,i)$ is the inner product of the representation of $u$ and $i$, which can be view as the similarity between $u$ and $i$.

\subsection{Localization of Point-of-interests}

The above pipeline follows a typical training framework for the matching stage. However, we argue that multi-modal recommendation is prone to be affected by noises within multiple modalities and the key for denoising lies in the localization of users' point-of-interests. We benefit from explainable machine learning \cite{Zhang_Wu_Zhu_2018,Selvaraju_Cogswell_Das_Vedantam_Parikh_Batra_2020} and propose to localize point-of-interests in the modality- and item- level based on gradients. Specifically, we obtain the gradient of the score $p_\theta'(i^+ \mid u)$ for positive item $i^+$ before the softmax, with respect to the item feature activations $\mathbf{i}_{t} \in \mathbb{R}^{d_i}$ and the modality feature activations $\mathbf{i}^m \in \mathbb{R}^{d_m}$, respectively. We apply global-average-pooling on the activations to obtain interest scores:
\begin{align}
	\alpha_{i_t} = \frac{1}{d_i} \sum_k^{d_i} \frac{\partial p_\theta'(i^+ \mid u)}{\partial \mathbf{i}_{t,k}}, \label{item}\\ 
	\alpha_{i_t^m} = \frac{1}{d_m} \sum_k^{d_m} \frac{\partial p_\theta'(i^+ \mid u)}{\partial \mathbf{i}^m_{t,k}} \label{modality}
\end{align}
In the image classification task, Grad-CAM \cite{Selvaraju_Cogswell_Das_Vedantam_Parikh_Batra_2020} uses the gradients of the target class prediction vector, with respect to feature map activation of a convolution layer to highlight the important regions within an image coarsely. Inspired by this, $\alpha_{i_t}$ and $\alpha_{i_t^m}$ could coarsely evaluate the interesting level of user $u$ on $i_t$ and $i_t^m$ when selecting the next interacted item. In section \ref{gradients}, we conduct an experiment to demonstrate the rationality of this method.

\subsection{User Augmentation}

We propose a novel way of learning user representations less sensitive to noisy modalities/items. We perform user augmentation in the modality-level and item-level based on the interest scores calculated by equations \ref{item} and \ref{modality}. Due to the page limits, we only formulate the generation of negative user embedding in details:   

\begin{figure}[t] \begin{center}
    \includegraphics[width=1\columnwidth]{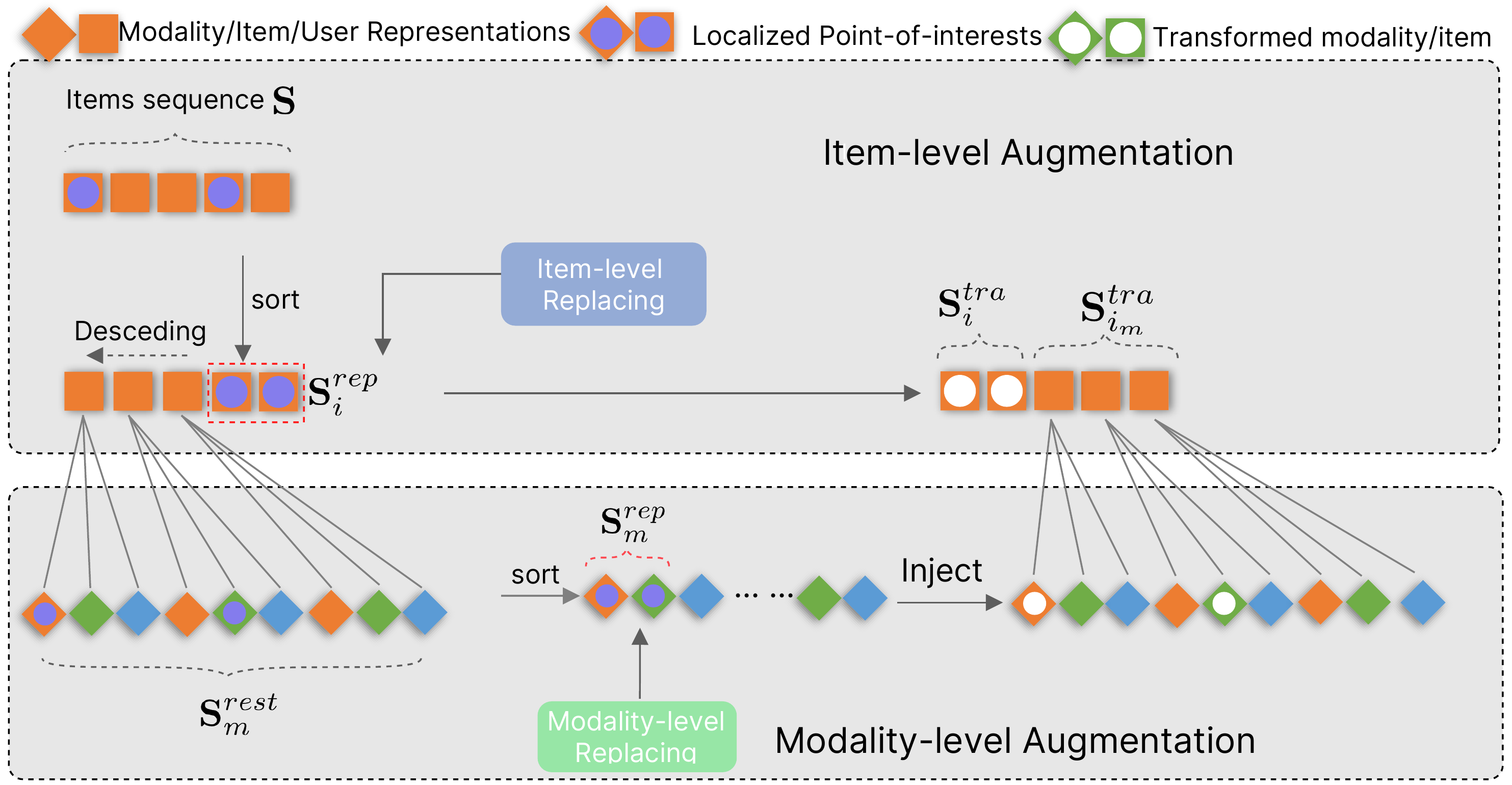}
    \caption{
    	The basic process of user augmentation. 
	}
\label{fig:user_aug}
\end{center} \end{figure} 
\vpara{Preliminary.} $\mathbf{S} = \{  \mathbf{i}_{u,t_1}, \mathbf{i}_{u,t_2}, \dots, \mathbf{i}_{u,t_{|\mathcal{S}|}} \}$ is the interaction item sequence of user $u$. For the sake of description convenience, we simplify it as $\mathbf{S} = \{ \mathbf{i}_{1}, \mathbf{i}_{2},\dots, \mathbf{i}_{l}, \dots, \mathbf{i}_{L} \} \in \mathbb{R}^{L \times d_i}$, where $\mathbf{i}_{l}$ is the item embedding vector. The items user $u$ do not interact with compose the set $\mathbf{U} \in \mathbb{R}^{N \times d_i}$ and their modalities set is $\mathbf{U}_m \in \mathbb{R}^{NM \times d_m}$, where M denotes the number of modality of an item.

\vpara{Item-level Augmentation.} We introduce a rate $\gamma_i$, which indicates the ratio of items transformed. As shown in Figure \ref{fig:user_aug}, given a user's behavior sequence $\mathbf{S}$, we sort it in descending order according to the interest scores computed by equation \ref{item}. We then choose the top $K_i$ items from the sorted set to constitute the replaced items set $\mathbf{S}^{rep}_i$ and calculate their similarity distribution with other items belonging to $\mathbf{U}$,

\begin{equation}
    \mathbf{S} = \{ \mathbf{i}_{1}, \mathbf{i}_{2}, \dots, \mathbf{i}_{L} \} \xrightarrow[select]{sort}
    \mathbf{S}^{rep}_i = \{ \mathbf{i}_{1}^{'}, \mathbf{i}_{2}^{'},\dots, \mathbf{i}_{K_i}^{'} \}
\end{equation}
\begin{equation}
    \mathbf{P}_i = \frac {\mathbf{D}_i - \mathbf{min}}{\mathbf{max} - \mathbf{min}},    \mathbf{D}_i = \mathbf{S}^{rep}_i \cdot \mathbf{U}^\intercal
\end{equation}

where  $\lfloor L \times \gamma_i \rfloor = K_i$, $\mathbf{S}^{rep}_i \in \mathbb{R}^{K_i \times d_i}$, $\mathbf{D}_i \in \mathbb{R}^{K_i \times N}$, $\mathbf{max} \in \mathbb{R}^{K_i}$ and $\mathbf{min} \in \mathbb{R}^{K_i}$ contains the maximum and minimum of each row in matrix $\mathbf{D}_i$. In order to select the dissimilar items from $\mathbf{U}$ to replace the items in $\mathbf{S}^{rep}_i$, We treat the distribution $1-\mathbf{P}_i$ as the probability of other items replacing the corresponding items. Based on the distribution $1-\mathbf{P}_i$, we obtain the transformed item set $\mathbf{S}^{tra}_i$.

\vpara{Modality-level Augmentation.} After item-level augmentation, the rest item set is,
\begin{equation}
    \mathbf{S}^{rest}_i = \mathbf{S} \setminus \mathbf{S}^{rep}_i 
\end{equation}
where $\mathbf{S}^{rest}_i \in \mathbb{R}^{K \times d_i}$, $K = L - K_i$. Its modality set denoted as $\mathbf{S}^{rest}_m = \{ \mathbf{i}_{1}^{1}, \mathbf{i}_{1}^{2}, \dots, \mathbf{i}_{1}^{m}, \dots, \mathbf{i}_{1}^{M}, \dots, \mathbf{i}_{K}^{1}, \mathbf{i}_{K}^{2}, \dots, \mathbf{i}_{K}^{m}, \dots, \mathbf{i}_{K}^{M} \} \in \mathbb{R}^{KM \times d_m}$, where $\mathbf{i}_{k}^m$ is the modality embedding vector. Similar to the item-level augmentation, we sort $\mathbf{S}^{rest}_m$ in descending order based on the modalities' interest scores computed by \ref{modality}. We then select the top $K_m$ modalities to be modified,

\begin{equation}
    \mathbf{S}^{rest}_m = \{ \mathbf{i}_{1}^1, \mathbf{i}_{1}^2, \dots, \mathbf{i}_{K}^M \} \xrightarrow[select]{sort}
    \mathbf{S}^{rep}_m = \{ \mathbf{i}_{1}^{m'}, \mathbf{i}_{2}^{m'},\dots, \mathbf{i}_{K_m}^{m'} \}
\end{equation}
\begin{equation}
    \mathbf{P}_m = \frac {\mathbf{D}_m - \mathbf{min}}{\mathbf{max} - \mathbf{min}},    \mathbf{D}_m = \mathbf{S}^{rep}_m \cdot \mathbf{U}_m^\intercal
\end{equation}

where  $\lfloor KM \times \gamma_m \rfloor = K_m$, $\mathbf{S}^{sep}_m \in \mathbb{R}^{K_m \times d_m}$, $\mathbf{D}_m \in \mathbb{R}^{K_m \times NM}$, $\mathbf{max} \in \mathbb{R}^{K_m}$ and $\mathbf{min} \in \mathbb{R}^{K_m}$ contains the maximum and minimum of each row in matrix $\mathbf{D}_m$. To select the dissimilar modalities from $\mathbf{U}_m$ to replace the modalities in $\mathbf{S}^{sep}_m$, We treat the distribution $1-\mathbf{P}_m$ as the probability of other modalities replacing the corresponding modalities. Based on the distribution $1-\mathbf{P}_m$, we generate the transformed modality set $\mathbf{S}^{tra}_m$.

\vpara{User Augmentation.} We inject the transformed modality set $\mathbf{S}^{tra}_m$ into $\mathbf{S}^{rest}_m$ and replace the corresponding original modalities. We then use the $\textbf{Modality-wise Encoder}$ to encode result set to obtain modality-level augmented items $\mathbf{S}^{tra}_{i_m}$. Finally, we use the $\textbf{Item-wise Encoder}$ to encode the transformed sequence, which is built by gathering the item-level and modality-level augmented items, \ie, $\mathbf{S}^{tra}_{i}$ and $\mathbf{S}^{tra}_{i_m}$, to obtain the negative user representation. We can also change the modalities and items with the top $\gamma_m$ and $\gamma_i$ \textbf{lowest} interest scores to generate the positive user representation. It is worth mentioning here that the similarity distributions used to select modalities/items do not need to subtract them from 1.

\subsection{Contrastive Learning}\label{contra}

With the augmented positive/negative user embeddings $\mathbf{u}^+$/$\mathbf{u}^-$, we conduct contrastive learning with the original user embedding $\mathbf{u}$. Mathemetically, we have:

\begin{align}
\label{eq:cont}
    \mathcal{L}_{cont} = - \log \frac{\sum_{j=1}^J  \exp \left(  \phi \left( \mathbf{u}^+_j, \mathbf{u} \right) \right)}{ \sum_{j=1}^J  \exp \left( \phi \left( \mathbf{u}^+_j, \mathbf{u} \right) \right) + \sum_{k=1}^{J} \exp \left( \phi \left( \mathbf{u}^-_k, \mathbf{u} \right) \right)} 
	 
\end{align}

where $J$ is the number of augmented positive/negative users. $\phi$ is a similarity function, and we employ dot product in the experiment. In a nutshell, this objective has the following advantages in the perspective of denoising multi-modal recommendation:
\begin{itemize}[leftmargin=*]
	\item In the augmented positive user sequence, noisy modalities/items are replaced by randomly sampled modality/item features. Pulling the user representation $\mathbf{u}$ and $\mathbf{u}^+$ together will lead to less overfitting on noisy modalities.
	\item Moreover, such a pulling makes the user encoder learns to trust more on point-of-interests, which are not transformed in the positive user sequence.
	\item Pushing the original user embedding $\mathbf{u}$ away from the negative user representation $\mathbf{u}^-$ will make the user encoder learn to be less affected by noisy modalities/items.
\end{itemize}
For augmented positive user embeddings, we also employ the recommendation-specific objective described in Section \ref{sec:oriobj}, further enhancing the user encoder to learn better representations. We denote the corresponding objective as $\mathcal{L}_{ssm}^{aug}$. Therefore, the final loss function for Demure can be written as:
\begin{align}
	\mathcal{L} = \mathcal{L}_{ssm} + \lambda_1 \mathcal{L}_{ssm}^{aug} + \lambda_2 \mathcal{L}_{cont}
\end{align}
where $\lambda_1$ and $\lambda_2$ are the hyperparameters to control the strength of each loss item.

\section{Experiments}

In this section, we conduct experiments on three publicly available datasets to reveal the effectiveness of Demure. We analyze Demure from the following perspectives:

\begin{itemize}[leftmargin=*]
	\item \textbf{RQ1}: How does Demure perform compared with the state-of-the-arts multi-modal recommendation models and other SOTA sequential recommenders on the multi-modal setting?
	\item \textbf{RQ2}: How do different designs (\eg, removing the contrastive learning objective, removing modality-/item level augmentation, different choices of transformation rate $\gamma_i / \gamma_m$) influence the performance of Demure?
	\item \textbf{RQ3}: Does Demure denoise the multi-modal modeling compared to the attentional base model?
\end{itemize}

\subsection{Experimental Settings} \label{sec:expset}
\vpara{Datasets.} Following \cite{Wei_Wang_Nie_He_Hong_Chua_2019,Wei_Wang_Nie_He_Chua_2020}, we conduct experiments on two publicly available multi-modal recommendation benchmarks, \ie, Tiktok and MovieLens \footnote{Kwai dataset used in their papers can not be downloaded in the official website https://www.kuaishou.com/activity/uimc.}. To evaluate Demure in a more broad domain, we incorporate the news recommendation dataset, MIND-NEWS \cite{Wu_Qiao_Chen_Wu_Qi_Lian_Liu_Xie_Gao_Wu_et_2020}, and build its multi-modal version MIND-MM by crawling the corresponding image for each news. Table \ref{tab:staData} lists the characteristics of these datasets.

\begin{itemize}[leftmargin=*]
	\item \textbf{Tiktok\footnote{http://ai-lab-challenge.bytedance.com/tce/vc/.}.} A micro-video recommendation dataset with each micro-video associated with pre-extracted video, textual, and acoustic features. Following \cite{Wei_Wang_Nie_He_Hong_Chua_2019}, we use a sampled version to speed up evaluation.
	\item \textbf{MovieLens\footnote{https://grouplens.org/datasets/movielens/.}.} To facilitate multi-modal recommendation evaluation, we choose the MovieLens-20M version which contains the video cover image, the video title/description, and categories. We adopt the pretrained glove word embedding \cite{Pennington_Socher_Manning_2014} of 100 dimension to represent each word in the title/description of a video. A pre-trained ResNet50 \footnote{\url{https://pytorch.org/vision/stable/models.html}} \cite{he2016deep} model is employed to extract a 2048-dimensional visual feature vector for the video cover image. We use a trainable embedding layer to map each category to a 64-dimensional feature vector.
	\item \textbf{MIND-MM.} MIND-NEWS \footnote{Original MIND dataset can be found in https://msnews.github.io/\#about-mind. We will release the multi-modal version to nourish the multi-modal recommendation research.} is a newly-released news recommendation that contains text and category solely. We augment the original dataset by crawling the images from the given news URLs. We disregard news that does not contain the image and use the remaining news to construct MIND-MM. We use the same pipeline as in the MovieLens dataset for feature extraction.
\end{itemize}

\begin{table}[!t]
\caption{Statistics of the Datasets. The V, A/C, T stands for Video, Audio/Category, Text modality respectively.}

\centering
\resizebox{\linewidth}{!}{
\begin{tabular}{l c c c c c c c}
\toprule
Dataset & \#Users & \#Items & \#Interactions & \#Density & V & A/C & T  \\
    \midrule
    Tiktok      & 40,357      & 89,335      & 1,208,213        & 0.033\%  & 128     & 128       & 128        	 \\ 
    MovieLens       & 138,493     & 25,381      & 19,882,894       & 0.56\%       & 2,048        &  64       &  100    	 \\
    MIND-NEWS       & 58,868      & 21,525      & 903,139     & 0.071\%      & 2,048     &  64       &  100        	 \\
    \bottomrule
\end{tabular}}%
    \label{tab:staData}
\end{table}
%

\begin{table*}[h]
\centering
    \caption{Performance Comparison across different multi-modal recommendation datasets and comparison methods. Base Model is the model modified from YoutubeDNN. Specifically, we extend the YoutubeDNN to the multi-modal setting and additionally utilize the attention mechanism instead of mean-pooling layer to aggregate the modalities/items into item/user representation.}
\begin{tabular}{l cc cc cc cc cc cc}
\toprule
&\multicolumn{4}{c}{ Tiktok } & \multicolumn{4}{c}{ MovieLens-20M } & \multicolumn{4}{c}{ MIND-MM } \\
\midrule
Model & \multicolumn{2}{c}{Metrics@20} & %
    \multicolumn{2}{c}{Metrics@50} & \multicolumn{2}{c}{Metrics@20} & %
    \multicolumn{2}{c}{Metrics@50} & \multicolumn{2}{c}{Metrics@20} & %
    \multicolumn{2}{c}{Metrics@50}\\
\cmidrule(lr){2-3}\cmidrule(lr){4-5}\cmidrule(lr){6-7}\cmidrule(lr){8-9}\cmidrule(lr){10-11}\cmidrule(lr){12-13}
    & Recall & NDCG & Recall & NDCG  & Recall & NDCG & Recall & NDCG & Recall & NDCG & Recall & NDCG \\
    \midrule
    YoutubeDNN       &0.025 	 &0.0492 	 &0.0451 	 &0.0864 	  &0.1251		 &0.3617 	 &0.249 	 &0.5432 	 &0.2057 	 &0.3373 	 &0.3226 	 &0.4851		\\ 
    GRU4Rec          &0.0266 	 &0.0577 	 &0.0447 	 &0.0897 	  &0.1255 		 &0.3734 	 &0.2368 	 &0.5391 	 &0.2014 	 &\underline{0.3469} 	 &0.323 	 &\underline{0.5069}  	\\ 
    MIND         	 &0.0254 	 &0.0493 	 &0.0494 	 &0.0947 	  &0.1308 		 &0.3655 	 &0.2449 	 &0.5366 	 &0.2063 	 &0.3393 	 &\underline{0.3244} 	 &0.4873  	\\ 
    ComiRec-DR           &0.0223 	 &0.0473 	 &0.0472 	 &0.0942 	  &0.1201 		 &0.3562 	 &0.2346 	 &0.5302 	 &0.1913 	 &0.3212 	 &0.3182 	 &0.4856          	\\ 
    ComiRec-SA 	 	 &\underline{0.0362} 	 &0.0708 	 &0.059 	 &0.1145 	  &0.1191 		 &\underline{0.3843} 	 &0.2436 	 &\underline{0.5726} 	 &0.2072 	 &0.3383 	 &0.316 	 &0.496 		\\ 
    MTIN           &0.0352 	 &\underline{0.0715} 	 &\underline{0.0697} 	 &\underline{0.1369} 	  &\underline{0.1328} 		 &0.3804 	 &\underline{0.2601} 	 &0.5631 	 &\underline{0.2092} 	 &0.3195 	 &0.3212 	 &0.4626  	\\ 
    \midrule
    Base Model 	 	 &0.026 	 &0.056 	 &0.0526 	 &0.1069 	  &0.1303 		 &0.373 	 &0.2572 	 &0.5581 	 &0.1906 	 &0.3237 	 &0.3115 	 &0.4795 	\\ 
    Demure           &  \textbf{0.0428	}  &  \textbf{0.0848	}  &  \textbf{0.0746	}  &  \textbf{0.1478	}  &  \textbf{0.1413	}  &  \textbf{0.4037	}  &  \textbf{0.2694	}  &  \textbf{0.5805	}  &  \textbf{0.2193	}  &  \textbf{0.3633	}  &  \textbf{0.3509	}  &  \textbf{0.5249	}  \\ 
        \midrule
    \%improv. 	 	 &18.23\% 	 &18.60\% 	 &7.03\% 	 &7.96\% 	  &6.4\% 		 &5.05\% 	 &3.58\% 	 &1.38\% 	 &4.83\% 	 &4.73\% 	 &8.17\% 	 &3.55\% 	\\ 
    \bottomrule
\end{tabular}
    \label{tab:modPer}
\end{table*}

\vpara{Evaluation Protocols.} We randomly split the users in the dataset into training, validation, and testing sets at the ratio 8:1:1. We note that splitting users can be more challenging compared to splitting user behavior sequences since models should confront out-of-distribution sequences, which can be essential for evaluating the denoising effectiveness. During testing, we view the first 80\% behaviors as historical behaviors and use the remaining 20\% behaviors for evaluation. Different from MMGCN that pairs each user with 1000 unobserved items that the user has not interacted with before, we propose to recall the positive items from the whole item gallery, which can be more realistic according to \cite{Krichene_Rendle_2020}.

\vpara{Baselines.} To evaluate the efficacy of Demure, we conduct a comparison with the following state-of-the-art sequential recommenders. As described in Section \ref{sec:problem}, we focus on the matching stage for recommendation and particularly choose matching comparison methods. We do not compare Demure with graph-based methods since they reply on the user-id embedding for collaborative filtering and cannot handle out-of-training users. We use two widely used numerical metrics, \ie, Recall@K and NDCG@K, and report the performance on top K=20 and 50 results recommended by all models on the test set.

\begin{itemize}[leftmargin=*]
	\item \textbf{YoutubeDNN \cite{Covington_Adams_Sargin_2016}}. YoutubeDNN combines historically interacted items using average-pooling. 
	\item \textbf{GRU4Rec \cite{Hidasi_Karatzoglou_Baltrunas_Tikk_2016}}. GRU4Rec replaces the naive average-pooling operation with a Gated Recurrent Unit \cite{Cho_Merrienboer_Gulcehre_Bahdanau_Bougares_Schwenk_Bengio_2014} to capture the sequential dependencies among items. 
	\item \textbf{MIND \cite{Li_Liu_Wu_Xu_Zhao_Huang_Kang_Chen_Li_Lee_2019}}. MIND uses the Capsule Network \cite{Sabour_Frosst_Hinton_2017} for extracting multiple interest vectors for a given behavior sequence. 
 	\item \textbf{ComiRec-DR \cite{Cen_Zhang_Zou_Zhou_Yang_Tang_2020}}. Besides the multi-interest representations for one user, ComiRec-DR further introduces a controllable factor to balance the recommendation quality and recommendation diversity. 
	\item \textbf{ComiRec-SA \cite{Cen_Zhang_Zou_Zhou_Yang_Tang_2020}}. ComiRec-SA differs from MIND by leveraging attention mechanisms for multi-interest extraction, which show empirical effectiveness. It further introduces a controllable factor to balance the recommendation quality and recommendation diversity. 
	\item \textbf{MTIN \cite{Jiang_Wang_Wei_Gao_Wang_Nie_2020}}. MTIN is a state-of-the-art content-based sequential recommender that leverages multi-scale time effects into user modeling and a group routing module to update interest vectors from the behavior sequence. 
\end{itemize}

\vpara{Implementation of Baselines.} We followed the implementations of the baselines in \url{https://github.com/THUDM/ComiRec} and extended them to a multi-modal version. Concretely, before encoding the item embeddings into user representation, we obtain the item embedding by aggregating modalities embeddings using attention mechanism.  

\vpara{Parameter Settings.} For all datasets, we train Demure with mini-batches of size 1024 and learning rate 0.001/0.003 for Tiktok/MovieLens and MIND-MM. We use Adam optimizer \cite{Kingma_Ba_2015} with $\beta_1=0.9$, $\beta_2=0.99$, and $1e-8$. We add L2 normalization for trainable parameters at a rate $1e-7$. We train for 20/20/100 epochs and keep at most 20/50/20 historical items for each user on the Tiktok/MovieLens/MIND-MMs dataset, respectively.

\subsection{Performance Comparison (RQ1)} We report the empirical results in Table \ref{tab:modPer}, where \%improv. indicates the relative improvement with the strongest state-of-the-art method which is highlighted in the underline. Analyzing the performance comparison, we have several key observations:
\begin{itemize}[leftmargin=*]
	\item Demure consistently outperforms all comparison models with a large margin in terms of all metrics. For example, the relative improvements over the strongest baselines are 18.23\%, 6.4\%, and 4.83\% in Tiktok, MovieLens-20M, and MIND-MM, \wrt the Recall@20 measurement. In addition, the base model without (w.o.) denoising is not a competitive one and cannot beat the strongest SOTA in most cases. We attribute the empirical effectiveness of Demure to: 1) by leveraging techniques in the explainable machine learning, Demure can effectively identify the point-of-interests/noises within the modality-level and the item-level. In contrast, most comparison methods disregard the noises introduced by multiple modalities and implicit feedback in user behaviors and thus suffer from noised user interest modeling. 2) the contrastive learning framework explicitly eliminates the effects of identified point-of-noises by pushing user representations obtained from noises and randomly sampled modalities/items away from the user representation used for prediction.
	\item Specifically, the conventional approach, YoutubeDNN, simply mean-pools the representations of all items and deeply suffers from the noises within multiple modalities and items, leading to inferior results in most cases. The update gate and reset gate in GRU4Rec might potentially alleviate the noises by differently treating multiple modalities/items. More advanced approaches like MIND, ComiRec-DR, ComiRec-SA construct multiple interest vectors using capsule network and attention mechanisms. These architectures might potentially aggregate point-of-interests/noises to part of the multi-interest vectors, and have less chance of being affected by the point-of-noises, and thus leading to performance improvement. The MTIN architecture implicitly deals with the implicit feedback problem, \ie, noises in the item-level, by computing a discount factor for each item based on the distance of its interest-item matching score and the average matching score of all items. This design helps MTIN achieves the best performance among all comparison methods, which further demonstrates the necessity of denoising recommendation. By explicitly alleviating noises in both the modality-level and item-level with explainable machine learning, Demure achieves the best results with substantial improvement. 
\end{itemize}

\subsection{Performance Comparison (RQ1)} We report the empirical results in Table \ref{tab:modPer}, where \%improv. indicates the relative improvement over the state-of-the-art method which is highlighted in the underline. Analyzing the performance comparison, we have several key observations:
\begin{itemize}[leftmargin=*]
	\item Demure consistently outperforms all comparison models with a large margin in terms of all metrics. For example, the relative improvements over the strongest baselines are 18.23\%, 6.4\%, and 4.83\% in Tiktok, MovieLens-20M, and MIND-MM, with respect to (\wrt) the Recall@20 measurement.
	\item Specifically, the conventional approach, YoutubeDNN, average-pools the representations of all items and deeply suffers from the noises within multiple modalities and items, leading to inferior results in most cases. More advanced approaches like MIND, ComiRec-DR, ComiRec-SA construct multiple interest vectors using capsule networks or attention mechanisms. These architectures might potentially aggregate point-of-interests/noises to part of the multi-interest vectors and have less chance of being affected by the point-of-noises, thus leading to performance improvement. The MTIN architecture implicitly deals with the implicit feedback problem, \ie, noises in the item-level, by computing a discount factor for each item based on the distance of its interest-item matching score and the average matching score of all items. This design helps MTIN achieve the best performance among all comparison methods, demonstrating the necessity of denoising recommendations. Demure achieves the best results with substantial improvement by explicitly alleviating noises in both the modality-level and item-level with explainable machine learning. 
	\item Demure is much more effective compared with the Base Model which is lack of user augmentation and contrastive learning objective. We attribute the empirical effectiveness of Demure to: \textbf{1)} by leveraging techniques in the explainable machine learning, Demure can effectively identify the point-of-interests/noises within the modality-level and the item-level. In contrast, most comparison methods disregard the noises introduced by multiple modalities and implicit feedback in user behaviors and thus suffer from noised user interest modeling; \textbf{2)} the contrastive learning framework explicitly eliminates the effects of identified noises by pushing user representations obtained from the augmented negative user sequence away from the original user representation just as we analyzed above in \ref{contra} section.
\end{itemize}

\begin{figure}[!t] \begin{center}
\includegraphics[width=\linewidth]{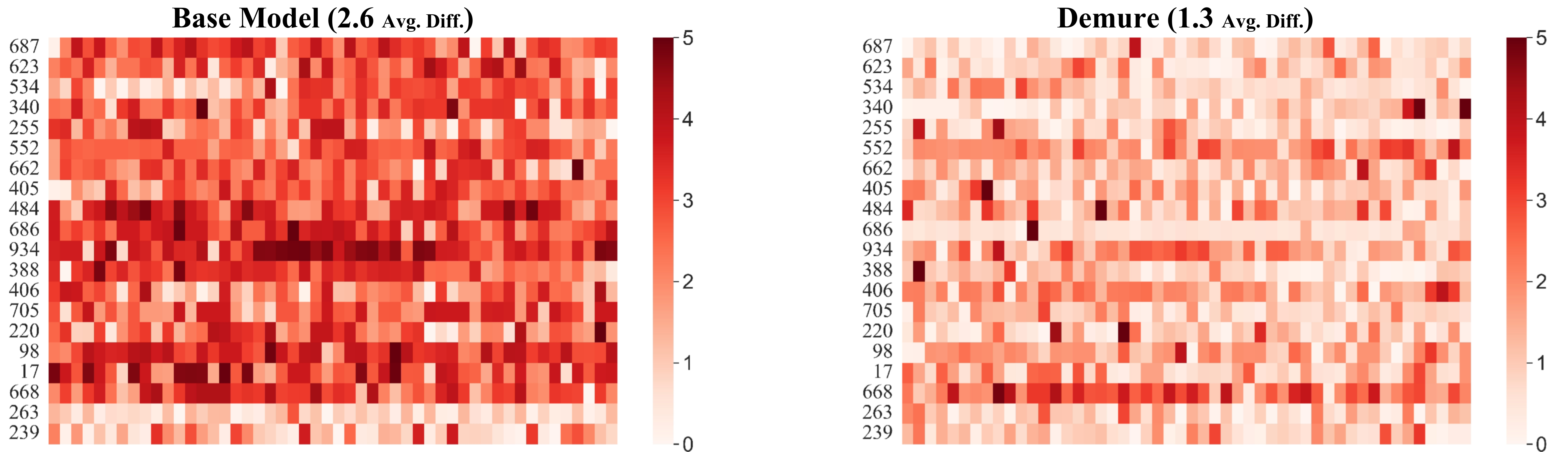}
\caption{
Visualization of the consistency between the predicted interest scores (gradients) and the corresponding ground-truth rating scores of users on items. The color strength of each pixel indicates the absolute difference. We plot maps of both the base model and the Demure.
    	}
\label{fig:gradients_map}
\end{center} \end{figure}

\subsection{Case Study (RQ2)}

\subsubsection{Visualization of the gradients.} \label{gradients}
It is hard to verify that treating the gradients as interest scores to localize the point-of-interests is reasonable without (\wo) explicit feedback in datasets. For that reason, we choose the ml-1m dataset to experiment, which includes ratings of users to the interacted items. We follow the same experimental setting of the above three datasets to train the Demure and Base Model with ml-1m. Then, we normalize the gradients within a user sequence and multiply it by 5 to keep the scale consistent with the actual rating. Considering the size and clarity of the figure, we randomly select 20 users and 50 items from their behavior sequence. We subtract their normalized and scaled gradients from actual rating scores and take the absolute values as difference scores. We then visualize the difference scores. The darker color means larger differences. We can see from Figure \ref{fig:gradients_map} that the distribution of the gradient of Demure after training is more tending to the actual ratings scores distribution compared with the Base Model. The experimental results can prove to some extent that items/modalities with higher gradients can be approximated as points-of-interests. The user augmentation and contrastive learning guide the localization of point-of-interests toward a more accurate direction.

\begin{figure}[!t] \begin{center}
\begin{subfigure}{.235\textwidth}
	\includegraphics[width=1\linewidth]{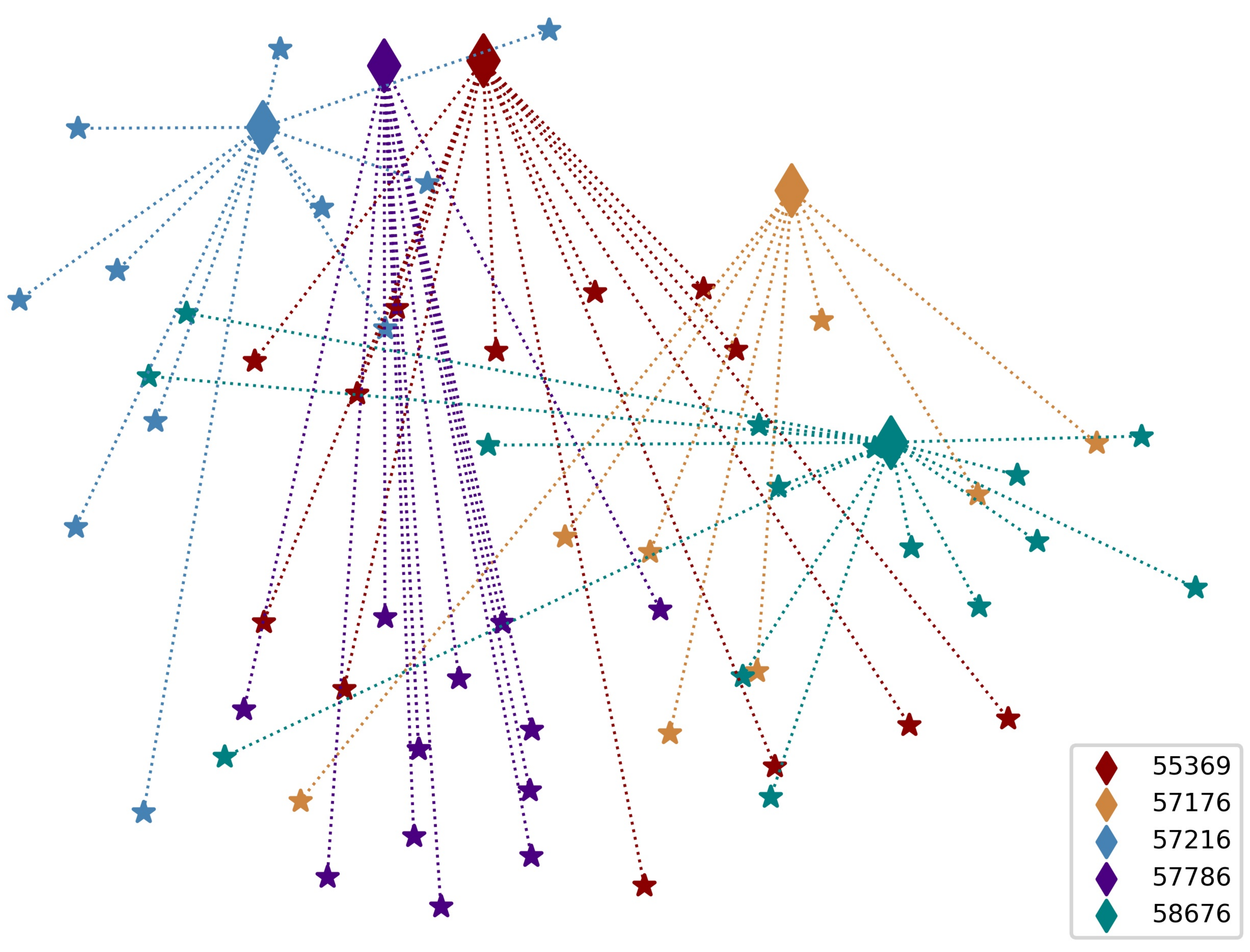}
    \caption{
    \footnotesize{w.o. denoising}
    	}
\label{fig:tsnebase}
\end{subfigure}
\begin{subfigure}{.235\textwidth}
	\includegraphics[width=1\linewidth]{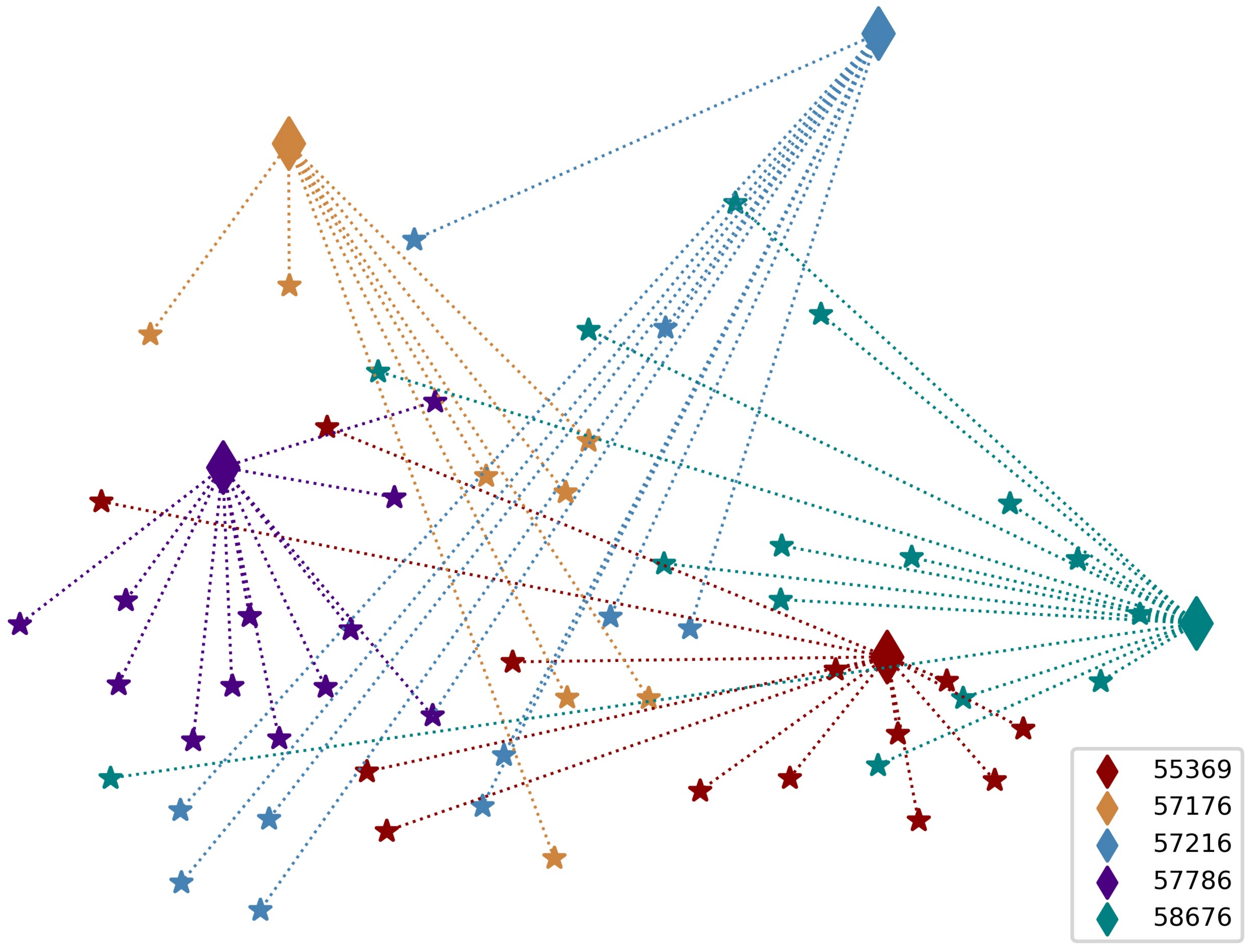}
    \caption{
   \footnotesize{Demure}
    	}
\label{fig:tsneDemure}
\end{subfigure}
    \caption{
     Visualization of the t-SNE transformed representations of randomly sampled users and their relevant items on the MIND-MM test set. These representations are obtained from multi-modal features. 
    	}
    \label{fig:tsne}
\end{center} \end{figure}

\subsubsection{Analysis of the embedding space.} In this section, we aim to analyze the denoising effect of the proposed framework on learning accurate user/item representations. We follow a pipeline proposed in \cite{Wang_He_Wang_Feng_Chua_2019} to perform the t-SNE transformation on the embeddings of randomly sampled five users as well as their relevant items in the MIND-MM test set. The results of the Base Model and Demure are plotted in Figure \ref{fig:tsnebase} and Figure \ref{fig:tsneDemure}, respectively. It is worth mentioning that the results are obtained after 5 runs. There are two observations:
\begin{itemize}[leftmargin=*]
	\item In Figure \ref{fig:tsneDemure}, items are discernibly clustered around the corresponding user (\eg, 57786) with less confusion compared to the model \wo denoising. This means that Demure learns more accurate user/item representations from multi-modal features that successfully capture point-of-interests during training at the modality-/item- level.
	\item Figure \ref{fig:tsnebase} shows an apparent entanglement between different users (\eg, 57216, 57786, 55369), leading to false-assignment of items and noised recommendation eventually. In contrast, users in Figure \ref{fig:tsneDemure} are more distinguishable from each other, demonstrating the effectiveness of denoising multi-modal recommendation for learning accurate user representations.
\end{itemize}

\subsubsection{Analysis of the attention weights.} 
\begin{figure}[!t] \begin{center}
\begin{subfigure}{.4\textwidth}
	\includegraphics[width=1\linewidth]{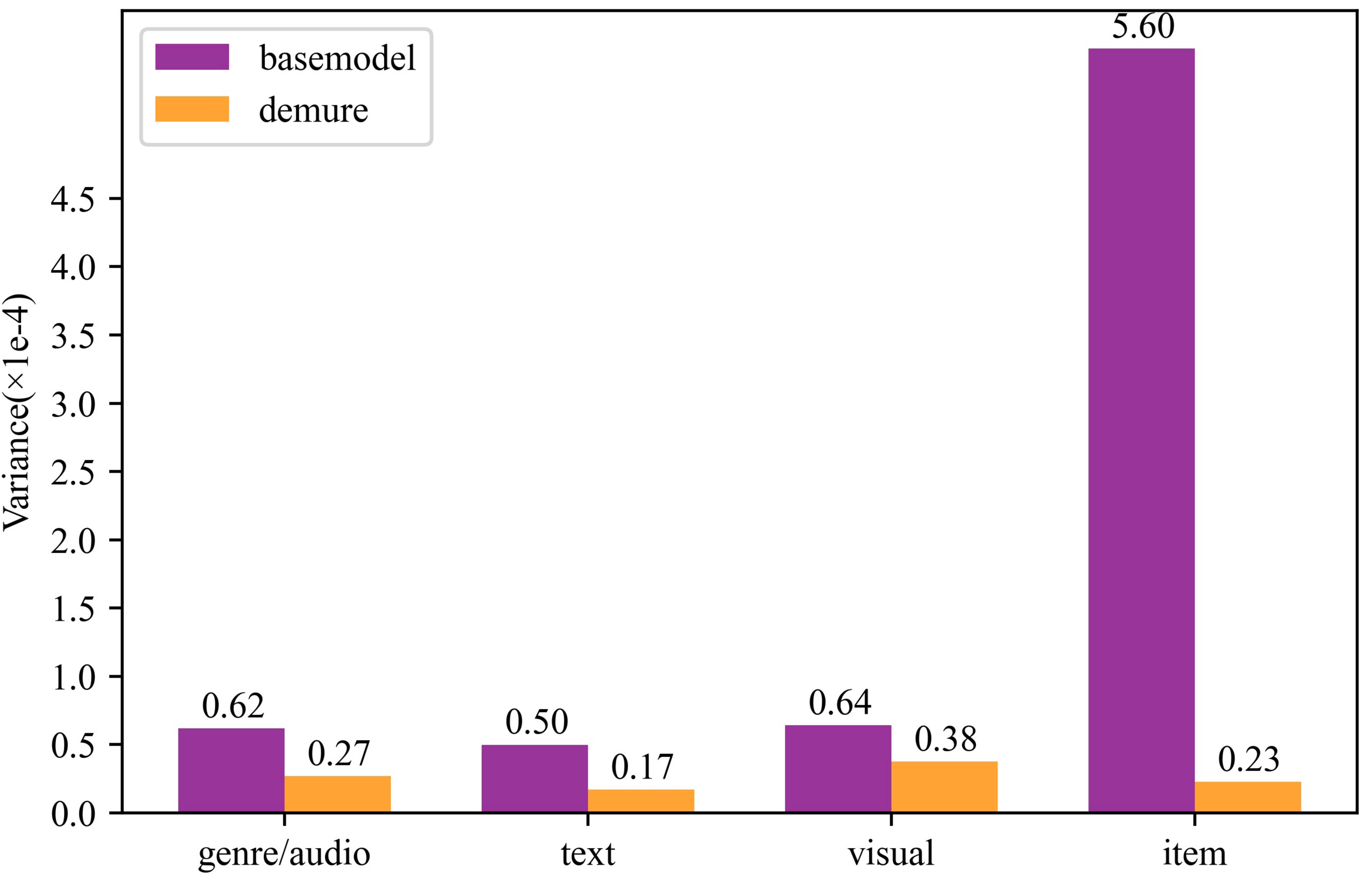}
    \caption{
    Variance of Item/Modality attention weights with respect to 10 \textit{\textbf{runs}}.
    	}
\label{fig:varRun}
\end{subfigure}
\begin{subfigure}{.4\textwidth}
	\includegraphics[width=1\linewidth]{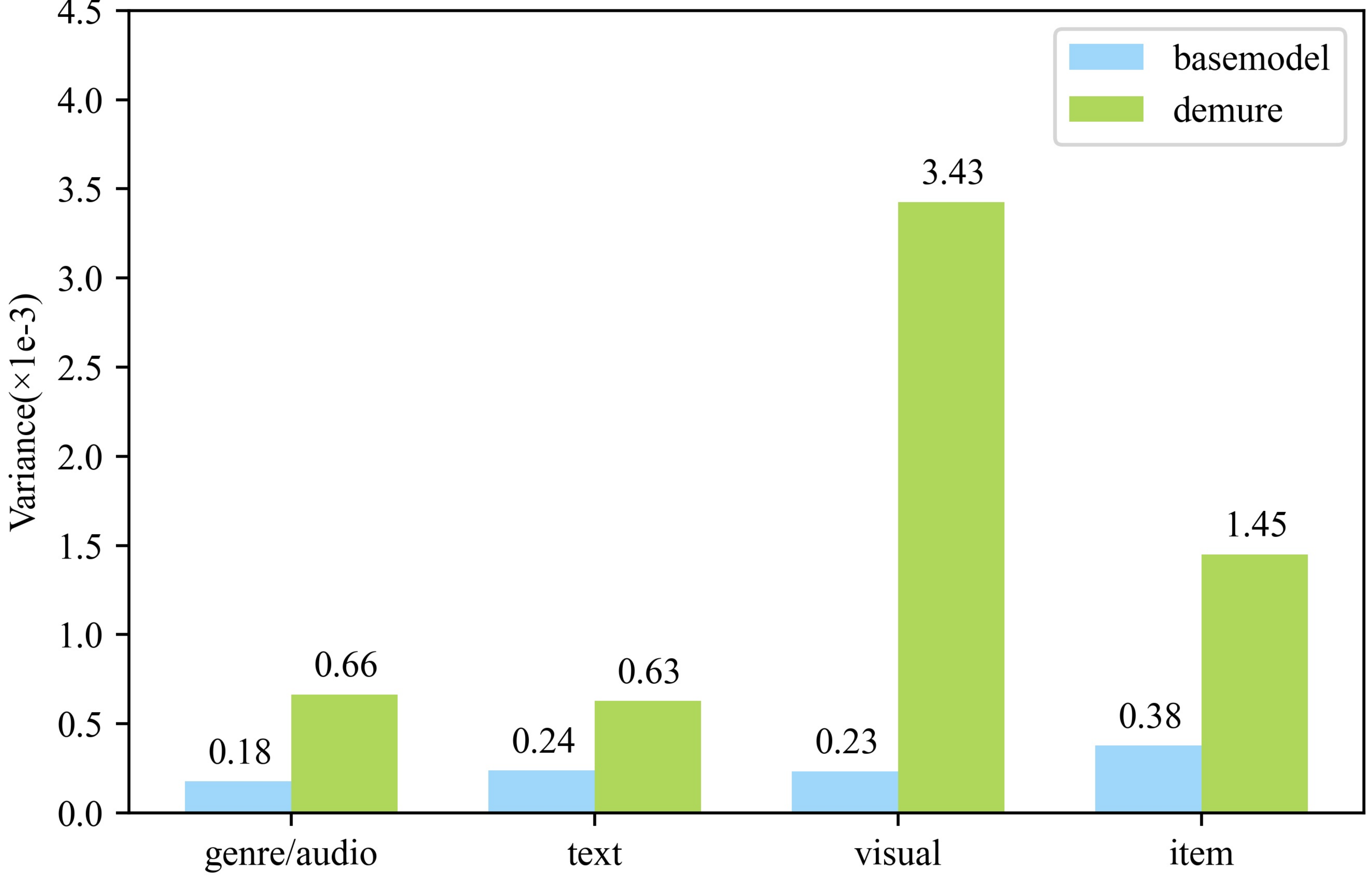}
    \caption{
   Variance of Item/Modality attention weights within each \textit{\textbf{user}}.
    	}
\label{fig:varUser}
\end{subfigure}
    \caption{
    Variance of Item/Modality attention weights across different scenarios.
    	}
    \label{fig:varUser}
\end{center} \end{figure}

The above visualization implicitly reflects the denoising effect on multi-modal features by depicting the correlations between items and users. To explicitly depict the correlations between users and different modalities, we propose to visualize the stability of learned attention weights of multiple modalities when representing the same user. Specifically, we run 10 experiments for the baseline and the Demure model until convergence and compute the modality-/item- level variances of attention weights with respect to different runs. We take the average variance of randomly sampled 1024 users, and the results are shown in Figure \ref{fig:varRun}. We observe significantly larger variances of the base model than those of Demure in both the modality-level and the item-level. Intuitively, noises in the data will lead to low confidence in essential assignments, which will eventually vary from time to time. These results potentially indicate that Demure can help denoise multi-modal recommendation and learn user representations that are aware of the point-of-interests/noises.

To shield the above analysis from the concern that Demure learns universally trivial attention weights (\eg, nearly the same attention weights for all modalities/items within one user), we compute the variances of attention weight within one user and take the average of 1024 users. The results are shown in Figure \ref{fig:varUser}. We can see that Demure assigns significantly different attention weights to different modalities/items within one user. Figure \ref{fig:varRun} and Figure \ref{fig:varUser} jointly demonstrate that Demure learns to identify point-of-interests/noises that are of high confidence.

\subsection{Quantitative Results (RQ3)}

\subsubsection{Study of augmentation rate $\gamma_i/\gamma_m$.}
The sum of augmentation rate $\gamma_i$ and $\gamma_m$ indicates the number of items being transformed in total. $\gamma_i$ and $\gamma_m$ reflect a tradeoff between item-level and fine-grained modality-level augmentation. We vary $\gamma_i + \gamma_m$, $\gamma_i$ and $\gamma_m$ to have an in-depth analysis on the effects of augmentation in different levels. The results are shown in Table \ref{tab:augrate}:
\begin{itemize}[leftmargin=*]
	\item Overall, the performance of Demure in all configurations shows improvement over the base model. In other words, Demure is insensitive to these hyper-parameters and can achieve improvement without bells and whistles. 
	\item The best results are achieved when $\gamma_i + \gamma_m = 0.6$. Too high or too low $\gamma_i + \gamma_m$ both achieve inferior results. Since $\gamma_i + \gamma_m$ indicates how many items are transformed (including item-level and modality-level transformation), transforming too many items ($\gamma_i + \gamma_m = 0.8$) might lead to collapse of positive samples, of which many point-of-interests might be transformed either, and transforming inadequate items ($\gamma_i + \gamma_m = 0.4$) might deteriorate the effectiveness of augmentation since augmented samples are mostly the same as the original. 
	\item In most cases, larger $\gamma_i$ will perform better than larger $\gamma_m$. For example, in $\gamma_i + \gamma_m=0.8 \text{ and } 0.6$, best results are mostly achieved by $\gamma_m=0.2$. Still, too low $\gamma_m$ (\eg, $\gamma_m=0.1$, $\gamma_i=0.3$) will lead to a performance drop.
	\item The results on three datasets of different domains (\ie, micro-video, video, news recommendation) have subtle differences, which means no universally optimal configuration for all application scenarios.

\end{itemize}

\begin{table}[h]
\centering
    \caption{Study of the effect of augmentation rates on the modality-level $\gamma_m$ and the item-level $\gamma_i$.}

\resizebox{\columnwidth}{22mm}{\begin{tabular}{c ccc cc cc }
\toprule
& & \multicolumn{2}{c}{ Tiktok } & \multicolumn{2}{c}{ MovieLens } & \multicolumn{2}{c}{ MIND-MM } \\
\midrule
   $\gamma_m + \gamma_i$ &  $\gamma_m$  & Recall & NDCG & Recall & NDCG  & Recall & NDCG  \\
    \midrule
    \multirow{3}{*}{$0.8$}  &   $0.6$      &0.0417 		 &0.0867 	 	&0.1404 	 &0.3991 	 	&0.2119 	 	&0.3569 	 \\
				  &   $0.4$  &0.04 	 	 &0.0797 	 	&0.1373 	 &0.3985 	 	&0.2102 	 	&0.3564 	 \\ 
        &   $0.2$  &0.0396 	 	 &0.0824 	 	&0.1411 	 &0.4000 	 	&0.2132 	 	&0.3559 	 \\ 
    \midrule
    \multirow{3}{*}{$0.6$}  &   $0.4$      &0.041 		 &0.0825 	 	&0.1385 	 &0.3935 	 	&0.2106 	 	&0.3537 	 \\
			  &   $0.3$  &0.0376 	 	 &0.0781 	 	&0.1406 	 &0.4005 	 	&0.2145 	 	&0.3591 	 \\ 
        &   $0.2$  &0.0428 	 	 &0.0848 	 	&0.1413 	 &0.4037 	 	&0.2193 	 	&0.3633 	 \\ 
    \midrule
    \multirow{3}{*}{$0.4$}  &   $0.3$      &0.0409		 &0.0842 	 	&0.1438 	 &0.4058 	 	&0.2097 	 	&0.3556 	 \\ 
			  &   $0.2$  &0.0387 	 	 &0.0789 	 	&0.1407 	 &0.3992 	 	&0.2164 	 	&0.3618 	 \\
        &   $0.1$  &0.0403 	 	 &0.082 	 	&0.1406 	 &0.3984 	 	&0.2126 	 	&0.3583 	 \\ 
    \bottomrule
\end{tabular}}

    \label{tab:augrate}
\end{table}

\subsubsection{Study of the number of augmented samples.} We evaluate the number of augmented positive/negative pairs, $J$. We vary $J$ among $\{1,2,4\}$. As illustrated in Table \ref{tab:numpairs}, we have following observations:
\begin{itemize}[leftmargin=*]
	\item The performance of Demure on $J=1$ and $J=2$ shows no significant differences across three datasets, which means that Demure is less sensitive to the number of augmented samples. We conduct the corresponding significant test, the p-value of Tiktok, MovieLens, MIND-MM is 0.153, 0.281, 0.397, respectively. All of them are larger than 0.05.  
	\item Further increasing $J$ will lead to a slight performance drop. We attribute this phenomenon to that simultaneously pulling the query user embedding closer to multiple positive samples and pushing it away from multiple negative ones might make it harder for optimization and eventually result in a trivial optimization. 
\end{itemize}

\begin{table}[h]
\centering
    \caption{Study of the number of augmented positive-negative user pairs.}
\resizebox{\columnwidth}{!}{\begin{tabular}{l cc cc cc }
\toprule
& \multicolumn{2}{c}{ Tiktok } & \multicolumn{2}{c}{ MovieLens } & \multicolumn{2}{c}{ MIND-MM } \\
\midrule
    & Recall & NDCG & Recall & NDCG  & Recall & NDCG  \\
    \midrule
    J = 1  	 & \textbf{0.0428} 	 	 & \textbf{0.0848} 	 	&0.1413 	 &0.4037 	 	& \textbf{0.2193} 	 	&0.3633 	 \\ 
    J = 2  	 &0.0393 	 	 &0.0813 	 	& \textbf{0.1432} 	 & \textbf{0.4046} 	 	&0.2172 	 	& \textbf{0.365} 	 \\
    J = 4  	 &0.0366 	 	 &0.0748 	 	&0.1398 	 &0.3985 	 	&0.2126 	 	&0.3572 	 \\
    \bottomrule
\end{tabular}}
    \label{tab:numpairs}
\end{table}

\section{Conclusion} \label{sec:conclusion}

In this paper, we propose to denoise multi-modal recommendation where introducing multiple modalities brings both useful features and noisy information. 
Towards this end, we devise a novel framework that explicitly localizes the point-of-interests and augments the original user behavior sequence in the modality-/item- level. Demure further encapsulates a contrastive learning objective that helps the user encoder learn preference representations that are more sensitive to point-of-interests and less sensitive to other modalities/items, thus denoising multi-modal recommendation. 
The experimental results demonstrate the effectiveness of Demure on two publicly available benchmarks and one multi-modal dataset built on a news recommendation dataset. Case studies qualitatively show that Demure helps denoise multi-modal recommendation.


\bibliographystyle{ACM-Reference-Format}
\bibliography{sections/Citation}


\begin{thebibliography}{51}


\ifx \showCODEN    \undefined \def \showCODEN     #1{\unskip}     \fi
\ifx \showDOI      \undefined \def \showDOI       #1{#1}\fi
\ifx \showISBNx    \undefined \def \showISBNx     #1{\unskip}     \fi
\ifx \showISBNxiii \undefined \def \showISBNxiii  #1{\unskip}     \fi
\ifx \showISSN     \undefined \def \showISSN      #1{\unskip}     \fi
\ifx \showLCCN     \undefined \def \showLCCN      #1{\unskip}     \fi
\ifx \shownote     \undefined \def \shownote      #1{#1}          \fi
\ifx \showarticletitle \undefined \def \showarticletitle #1{#1}   \fi
\ifx \showURL      \undefined \def \showURL       {\relax}        \fi
\providecommand\bibfield[2]{#2}
\providecommand\bibinfo[2]{#2}
\providecommand\natexlab[1]{#1}
\providecommand\showeprint[2][]{arXiv:#2}

\bibitem[Wan(2020)]%
        {Wang2020_Chorus}
 \bibinfo{year}{2020}\natexlab{}.
\newblock \showarticletitle{{Make It a Chorus: Knowledge-and Time-aware Item
  Modeling for Sequential Recommendation}}.
\newblock \bibinfo{journal}{\emph{SIGIR 2020 - Proceedings of the 43rd
  International ACM SIGIR Conference on Research and Development in Information
  Retrieval}} (\bibinfo{year}{2020}), \bibinfo{pages}{109--118}.
\newblock
\showISBNx{9781450380164}
\urldef\tempurl%
\url{https://doi.org/10.1145/3397271.3401131}
\showDOI{\tempurl}


\bibitem[Cen et~al\mbox{.}(2020)]%
        {Cen_Zhang_Zou_Zhou_Yang_Tang_2020}
\bibfield{author}{\bibinfo{person}{Yukuo Cen}, \bibinfo{person}{Jianwei Zhang},
  \bibinfo{person}{Xu Zou}, \bibinfo{person}{Chang Zhou},
  \bibinfo{person}{Hongxia Yang}, {and} \bibinfo{person}{Jie Tang}.}
  \bibinfo{year}{2020}\natexlab{}.
\newblock \showarticletitle{Controllable Multi-Interest Framework for
  Recommendation.}. In \bibinfo{booktitle}{\emph{KDD ’20: The 26th ACM SIGKDD
  Conference on Knowledge Discovery and Data Mining, Virtual Event, CA, USA,
  August 23-27, 2020}}.
\newblock


\bibitem[Chelliah et~al\mbox{.}(2019)]%
        {Chelliah_Biswas_Dhakad_2019}
\bibfield{author}{\bibinfo{person}{Muthusamy Chelliah}, \bibinfo{person}{Soma
  Biswas}, {and} \bibinfo{person}{Lucky Dhakad}.}
  \bibinfo{year}{2019}\natexlab{}.
\newblock \showarticletitle{Principle-to-program: Neural Fashion Recommendation
  with Multi-modal Input.}. In \bibinfo{booktitle}{\emph{Proceedings of the
  27th ACM International Conference on Multimedia, MM 2019, Nice, France,
  October 21-25, 2019}}.
\newblock


\bibitem[Chen and He(2017)]%
        {Chen2017}
\bibfield{author}{\bibinfo{person}{Jingyuan Chen} {and}
  \bibinfo{person}{Xiangnan He}.} \bibinfo{year}{2017}\natexlab{}.
\newblock \showarticletitle{{A entive Collaborative Filtering : Multimedia
  Recommendation with Item- and Component-Level A ention}}.
\newblock  (\bibinfo{year}{2017}), \bibinfo{pages}{335--344}.
\newblock
\showISBNx{9781450350228}


\bibitem[Chen et~al\mbox{.}(2017)]%
        {Chen_Zhang_He_Nie_Liu_Chua_2017}
\bibfield{author}{\bibinfo{person}{Jingyuan Chen}, \bibinfo{person}{Hanwang
  Zhang}, \bibinfo{person}{Xiangnan He}, \bibinfo{person}{Liqiang Nie},
  \bibinfo{person}{Wei Liu}, {and} \bibinfo{person}{Tat-Seng Chua}.}
  \bibinfo{year}{2017}\natexlab{}.
\newblock \showarticletitle{Attentive Collaborative Filtering: Multimedia
  Recommendation with Item- and Component-Level Attention.}. In
  \bibinfo{booktitle}{\emph{Proceedings of the 40th International ACM SIGIR
  Conference on Research and Development in Information Retrieval, Shinjuku,
  Tokyo, Japan, August 7-11, 2017}}.
\newblock


\bibitem[Chen et~al\mbox{.}(2020)]%
        {Chen_Kornblith_Norouzi_Hinton_2020}
\bibfield{author}{\bibinfo{person}{Ting Chen}, \bibinfo{person}{Simon
  Kornblith}, \bibinfo{person}{Mohammad Norouzi}, {and}
  \bibinfo{person}{Geoffrey~E. Hinton}.} \bibinfo{year}{2020}\natexlab{}.
\newblock \showarticletitle{A Simple Framework for Contrastive Learning of
  Visual Representations.}. In \bibinfo{booktitle}{\emph{Proceedings of the
  37th International Conference on Machine Learning, ICML 2020, 13-18 July
  2020, Virtual Event.}}
\newblock


\bibitem[Cho et~al\mbox{.}(2014)]%
        {Cho_Merrienboer_Gulcehre_Bahdanau_Bougares_Schwenk_Bengio_2014}
\bibfield{author}{\bibinfo{person}{Kyunghyun Cho}, \bibinfo{person}{Bart~van
  Merrienboer}, \bibinfo{person}{Caglar Gulcehre}, \bibinfo{person}{Dzmitry
  Bahdanau}, \bibinfo{person}{Fethi Bougares}, \bibinfo{person}{Holger
  Schwenk}, {and} \bibinfo{person}{Yoshua Bengio}.}
  \bibinfo{year}{2014}\natexlab{}.
\newblock \showarticletitle{Learning Phrase Representations using RNN
  Encoder-Decoder for Statistical Machine Translation.}. In
  \bibinfo{booktitle}{\emph{Proceedings of the 2014 Conference on Empirical
  Methods in Natural Language Processing, EMNLP 2014, October 25-29, 2014,
  Doha, Qatar, A meeting of SIGDAT, a Special Interest Group of the ACL}}.
\newblock


\bibitem[Covington et~al\mbox{.}(2016)]%
        {Covington_Adams_Sargin_2016}
\bibfield{author}{\bibinfo{person}{Paul Covington}, \bibinfo{person}{Jay
  Adams}, {and} \bibinfo{person}{Emre Sargin}.}
  \bibinfo{year}{2016}\natexlab{}.
\newblock \showarticletitle{Deep Neural Networks for YouTube Recommendations.}.
  In \bibinfo{booktitle}{\emph{Proceedings of the 10th ACM Conference on
  Recommender Systems, Boston, MA, USA, September 15-19, 2016}}.
\newblock


\bibitem[Du et~al\mbox{.}(2020)]%
        {Du_Wang_He_Li_Tang_Chua_2020}
\bibfield{author}{\bibinfo{person}{Xiaoyu Du}, \bibinfo{person}{Xiang Wang},
  \bibinfo{person}{Xiangnan He}, \bibinfo{person}{Zechao Li},
  \bibinfo{person}{Jinhui Tang}, {and} \bibinfo{person}{Tat-Seng Chua}.}
  \bibinfo{year}{2020}\natexlab{}.
\newblock \showarticletitle{How to Learn Item Representation for Cold-Start
  Multimedia Recommendation?}. In \bibinfo{booktitle}{\emph{MM ’20: The 28th
  ACM International Conference on Multimedia, Virtual Event / Seattle, WA, USA,
  October 12-16, 2020}}.
\newblock


\bibitem[Fu et~al\mbox{.}(2019)]%
        {Fu2019}
\bibfield{author}{\bibinfo{person}{Wenjing Fu}, \bibinfo{person}{Zhaohui Peng},
  \bibinfo{person}{Senzhang Wang}, \bibinfo{person}{Yang Xu}, {and}
  \bibinfo{person}{Jin Li}.} \bibinfo{year}{2019}\natexlab{}.
\newblock \showarticletitle{{Deeply fusing reviews and contents for cold start
  users in cross-domain recommendation systems}}.
\newblock \bibinfo{journal}{\emph{33rd AAAI Conference on Artificial
  Intelligence, AAAI 2019, 31st Innovative Applications of Artificial
  Intelligence Conference, IAAI 2019 and the 9th AAAI Symposium on Educational
  Advances in Artificial Intelligence, EAAI 2019}} (\bibinfo{year}{2019}),
  \bibinfo{pages}{94--101}.
\newblock
\showISBNx{9781577358091}
\showISSN{2159-5399}
\urldef\tempurl%
\url{https://doi.org/10.1609/aaai.v33i01.330194}
\showDOI{\tempurl}


\bibitem[He et~al\mbox{.}(2020)]%
        {He_Fan_Wu_Xie_Girshick_2020}
\bibfield{author}{\bibinfo{person}{Kaiming He}, \bibinfo{person}{Haoqi Fan},
  \bibinfo{person}{Yuxin Wu}, \bibinfo{person}{Saining Xie}, {and}
  \bibinfo{person}{Ross~B. Girshick}.} \bibinfo{year}{2020}\natexlab{}.
\newblock \showarticletitle{Momentum Contrast for Unsupervised Visual
  Representation Learning.}. In \bibinfo{booktitle}{\emph{2020 IEEE/CVF
  Conference on Computer Vision and Pattern Recognition, CVPR 2020, Seattle,
  WA, USA, June 13-19, 2020}}.
\newblock


\bibitem[He et~al\mbox{.}(2016)]%
        {he2016deep}
\bibfield{author}{\bibinfo{person}{Kaiming He}, \bibinfo{person}{Xiangyu
  Zhang}, \bibinfo{person}{Shaoqing Ren}, {and} \bibinfo{person}{Jian Sun}.}
  \bibinfo{year}{2016}\natexlab{}.
\newblock \showarticletitle{Deep residual learning for image recognition}. In
  \bibinfo{booktitle}{\emph{Proceedings of the IEEE conference on computer
  vision and pattern recognition}}. \bibinfo{pages}{770--778}.
\newblock


\bibitem[Hidasi et~al\mbox{.}(2016)]%
        {Hidasi_Karatzoglou_Baltrunas_Tikk_2016}
\bibfield{author}{\bibinfo{person}{Balázs Hidasi}, \bibinfo{person}{Alexandros
  Karatzoglou}, \bibinfo{person}{Linas Baltrunas}, {and}
  \bibinfo{person}{Domonkos Tikk}.} \bibinfo{year}{2016}\natexlab{}.
\newblock \showarticletitle{Session-based Recommendations with Recurrent Neural
  Networks.}. In \bibinfo{booktitle}{\emph{4th International Conference on
  Learning Representations, ICLR 2016, San Juan, Puerto Rico, May 2-4, 2016,
  Conference Track Proceedings}}.
\newblock


\bibitem[Huang et~al\mbox{.}(2019)]%
        {Huang_Fang_Qian_Sang_Li_Xu_2019}
\bibfield{author}{\bibinfo{person}{Xiaowen Huang}, \bibinfo{person}{Quan Fang},
  \bibinfo{person}{Shengsheng Qian}, \bibinfo{person}{Jitao Sang},
  \bibinfo{person}{Yan Li}, {and} \bibinfo{person}{Changsheng Xu}.}
  \bibinfo{year}{2019}\natexlab{}.
\newblock \showarticletitle{Explainable Interaction-driven User Modeling over
  Knowledge Graph for Sequential Recommendation.}. In
  \bibinfo{booktitle}{\emph{Proceedings of the 27th ACM International
  Conference on Multimedia, MM 2019, Nice, France, October 21-25, 2019}}.
\newblock


\bibitem[Jiang et~al\mbox{.}(2019)]%
        {Jiang_Wang_Liu_Nie_Duan_Xu_2019}
\bibfield{author}{\bibinfo{person}{Hao Jiang}, \bibinfo{person}{Wenjie Wang},
  \bibinfo{person}{Meng Liu}, \bibinfo{person}{Liqiang Nie},
  \bibinfo{person}{Ling-Yu Duan}, {and} \bibinfo{person}{Changsheng Xu}.}
  \bibinfo{year}{2019}\natexlab{}.
\newblock \showarticletitle{Market2Dish: A Health-aware Food Recommendation
  System.}. In \bibinfo{booktitle}{\emph{Proceedings of the 27th ACM
  International Conference on Multimedia, MM 2019, Nice, France, October 21-25,
  2019}}.
\newblock


\bibitem[Jiang et~al\mbox{.}(2020)]%
        {Jiang_Wang_Wei_Gao_Wang_Nie_2020}
\bibfield{author}{\bibinfo{person}{Hao Jiang}, \bibinfo{person}{Wenjie Wang},
  \bibinfo{person}{Yinwei Wei}, \bibinfo{person}{Zan Gao},
  \bibinfo{person}{Yinglong Wang}, {and} \bibinfo{person}{Liqiang Nie}.}
  \bibinfo{year}{2020}\natexlab{}.
\newblock \showarticletitle{What Aspect Do You Like: Multi-scale Time-aware
  User Interest Modeling for Micro-video Recommendation.}. In
  \bibinfo{booktitle}{\emph{MM ’20: The 28th ACM International Conference on
  Multimedia, Virtual Event / Seattle, WA, USA, October 12-16, 2020}}.
\newblock


\bibitem[Kang and Park(2020)]%
        {Kang_Park_2020}
\bibfield{author}{\bibinfo{person}{Minguk Kang} {and} \bibinfo{person}{Jaesik
  Park}.} \bibinfo{year}{2020}\natexlab{}.
\newblock \showarticletitle{ContraGAN: Contrastive Learning for Conditional
  Image Generation.}. In \bibinfo{booktitle}{\emph{Advances in Neural
  Information Processing Systems 33: Annual Conference on Neural Information
  Processing Systems 2020, NeurIPS 2020, December 6-12, 2020, virtual.}}
\newblock


\bibitem[Kingma and Ba(2015)]%
        {Kingma_Ba_2015}
\bibfield{author}{\bibinfo{person}{Diederik~P. Kingma} {and}
  \bibinfo{person}{Jimmy Ba}.} \bibinfo{year}{2015}\natexlab{}.
\newblock \showarticletitle{Adam: A Method for Stochastic Optimization.}. In
  \bibinfo{booktitle}{\emph{3rd International Conference on Learning
  Representations, ICLR 2015, San Diego, CA, USA, May 7-9, 2015, Conference
  Track Proceedings}}.
\newblock


\bibitem[Krichene and Rendle(2020)]%
        {Krichene_Rendle_2020}
\bibfield{author}{\bibinfo{person}{Walid Krichene} {and}
  \bibinfo{person}{Steffen Rendle}.} \bibinfo{year}{2020}\natexlab{}.
\newblock \showarticletitle{On Sampled Metrics for Item Recommendation.}. In
  \bibinfo{booktitle}{\emph{KDD ’20: The 26th ACM SIGKDD Conference on
  Knowledge Discovery and Data Mining, Virtual Event, CA, USA, August 23-27,
  2020}}.
\newblock


\bibitem[Laskin et~al\mbox{.}(2020)]%
        {Laskin_Srinivas_Abbeel_2020}
\bibfield{author}{\bibinfo{person}{Michael Laskin}, \bibinfo{person}{Aravind
  Srinivas}, {and} \bibinfo{person}{Pieter Abbeel}.}
  \bibinfo{year}{2020}\natexlab{}.
\newblock \showarticletitle{CURL: Contrastive Unsupervised Representations for
  Reinforcement Learning.}. In \bibinfo{booktitle}{\emph{Proceedings of the
  37th International Conference on Machine Learning, ICML 2020, 13-18 July
  2020, Virtual Event.}}
\newblock


\bibitem[Li et~al\mbox{.}(2019a)]%
        {Li_Liu_Wu_Xu_Zhao_Huang_Kang_Chen_Li_Lee_2019}
\bibfield{author}{\bibinfo{person}{Chao Li}, \bibinfo{person}{Zhiyuan Liu},
  \bibinfo{person}{Mengmeng Wu}, \bibinfo{person}{Yuchi Xu},
  \bibinfo{person}{Huan Zhao}, \bibinfo{person}{Pipei Huang},
  \bibinfo{person}{Guoliang Kang}, \bibinfo{person}{Qiwei Chen},
  \bibinfo{person}{Wei Li}, {and} \bibinfo{person}{Dik~Lun Lee}.}
  \bibinfo{year}{2019}\natexlab{a}.
\newblock \showarticletitle{Multi-Interest Network with Dynamic Routing for
  Recommendation at Tmall.}. In \bibinfo{booktitle}{\emph{Proceedings of the
  28th ACM International Conference on Information and Knowledge Management,
  CIKM 2019, Beijing, China, November 3-7, 2019.}}
\newblock


\bibitem[Li et~al\mbox{.}(2019b)]%
        {Li_Liu_Yin_Cui_Xu_Nie_2019}
\bibfield{author}{\bibinfo{person}{Yongqi Li}, \bibinfo{person}{Meng Liu},
  \bibinfo{person}{Jianhua Yin}, \bibinfo{person}{Chaoran Cui},
  \bibinfo{person}{Xin-Shun Xu}, {and} \bibinfo{person}{Liqiang Nie}.}
  \bibinfo{year}{2019}\natexlab{b}.
\newblock \showarticletitle{Routing Micro-videos via A Temporal Graph-guided
  Recommendation System.}. In \bibinfo{booktitle}{\emph{Proceedings of the 27th
  ACM International Conference on Multimedia, MM 2019, Nice, France, October
  21-25, 2019}}.
\newblock


\bibitem[Lu et~al\mbox{.}(2019)]%
        {Lu2019}
\bibfield{author}{\bibinfo{person}{Jiasen Lu}, \bibinfo{person}{Dhruv Batra},
  \bibinfo{person}{Devi Parikh}, {and} \bibinfo{person}{Stefan Lee}.}
  \bibinfo{year}{2019}\natexlab{}.
\newblock \showarticletitle{{ViLBERT: Pretraining task-agnostic visiolinguistic
  representations for vision-and-language tasks}}.
\newblock \bibinfo{journal}{\emph{Advances in Neural Information Processing
  Systems}}  \bibinfo{volume}{32} (\bibinfo{year}{2019}),
  \bibinfo{pages}{1--11}.
\newblock
\showISSN{10495258}
\showeprint[arxiv]{1908.02265}


\bibitem[Meng et~al\mbox{.}(2020)]%
        {Meng_Feng_He_Gao_Chua_2020}
\bibfield{author}{\bibinfo{person}{Lei Meng}, \bibinfo{person}{Fuli Feng},
  \bibinfo{person}{Xiangnan He}, \bibinfo{person}{Xiaoyan Gao}, {and}
  \bibinfo{person}{Tat-Seng Chua}.} \bibinfo{year}{2020}\natexlab{}.
\newblock \showarticletitle{Heterogeneous Fusion of Semantic and Collaborative
  Information for Visually-Aware Food Recommendation.}. In
  \bibinfo{booktitle}{\emph{MM ’20: The 28th ACM International Conference on
  Multimedia, Virtual Event / Seattle, WA, USA, October 12-16, 2020}}.
\newblock


\bibitem[Pennington et~al\mbox{.}(2014)]%
        {Pennington_Socher_Manning_2014}
\bibfield{author}{\bibinfo{person}{Jeffrey Pennington},
  \bibinfo{person}{Richard Socher}, {and} \bibinfo{person}{Christopher~D.
  Manning}.} \bibinfo{year}{2014}\natexlab{}.
\newblock \showarticletitle{Glove: Global Vectors for Word Representation.}. In
  \bibinfo{booktitle}{\emph{Proceedings of the 2014 Conference on Empirical
  Methods in Natural Language Processing, EMNLP 2014, October 25-29, 2014,
  Doha, Qatar, A meeting of SIGDAT, a Special Interest Group of the ACL}}.
\newblock


\bibitem[Qi et~al\mbox{.}(2021)]%
        {Qi2021}
\bibfield{author}{\bibinfo{person}{Tao Qi}, \bibinfo{person}{Fangzhao Wu},
  \bibinfo{person}{Chuhan Wu}, {and} \bibinfo{person}{Yongfeng Huang}.}
  \bibinfo{year}{2021}\natexlab{}.
\newblock \showarticletitle{{Personalized News Recommendation with
  Knowledge-aware Interactive Matching}}.
\newblock  (\bibinfo{year}{2021}).
\newblock
\showISBNx{9781450380379}
\showeprint[arxiv]{arXiv:2104.10083v3}


\bibitem[Ren et~al\mbox{.}(2020)]%
        {Ren2020}
\bibfield{author}{\bibinfo{person}{Ruiyang Ren}, \bibinfo{person}{Zhaoyang
  Liu}, \bibinfo{person}{Yaliang Li}, \bibinfo{person}{Wayne~Xin Zhao},
  \bibinfo{person}{Hui Wang}, \bibinfo{person}{Bolin Ding}, {and}
  \bibinfo{person}{Ji~Rong Wen}.} \bibinfo{year}{2020}\natexlab{}.
\newblock \showarticletitle{{Sequential Recommendation with Self-Attentive
  Multi-Adversarial Network}}.
\newblock \bibinfo{journal}{\emph{SIGIR 2020 - Proceedings of the 43rd
  International ACM SIGIR Conference on Research and Development in Information
  Retrieval}} (\bibinfo{year}{2020}), \bibinfo{pages}{89--98}.
\newblock
\showISBNx{9781450380164}
\urldef\tempurl%
\url{https://doi.org/10.1145/3397271.3401111}
\showDOI{\tempurl}
\showeprint[arxiv]{2005.10602}


\bibitem[Rosenbaum and Rubin(1983)]%
        {rosenbaum1983central}
\bibfield{author}{\bibinfo{person}{Paul~R Rosenbaum} {and}
  \bibinfo{person}{Donald~B Rubin}.} \bibinfo{year}{1983}\natexlab{}.
\newblock \showarticletitle{The central role of the propensity score in
  observational studies for causal effects}.
\newblock \bibinfo{journal}{\emph{Biometrika}} \bibinfo{volume}{70},
  \bibinfo{number}{1} (\bibinfo{year}{1983}), \bibinfo{pages}{41--55}.
\newblock


\bibitem[Sabour et~al\mbox{.}(2017)]%
        {Sabour_Frosst_Hinton_2017}
\bibfield{author}{\bibinfo{person}{Sara Sabour}, \bibinfo{person}{Nicholas
  Frosst}, {and} \bibinfo{person}{Geoffrey~E. Hinton}.}
  \bibinfo{year}{2017}\natexlab{}.
\newblock \showarticletitle{Dynamic Routing Between Capsules.}. In
  \bibinfo{booktitle}{\emph{Advances in Neural Information Processing Systems
  30: Annual Conference on Neural Information Processing Systems 2017, December
  4-9, 2017, Long Beach, CA, USA}}.
\newblock


\bibitem[Selvaraju et~al\mbox{.}(2020)]%
        {Selvaraju_Cogswell_Das_Vedantam_Parikh_Batra_2020}
\bibfield{author}{\bibinfo{person}{Ramprasaath~R. Selvaraju},
  \bibinfo{person}{Michael Cogswell}, \bibinfo{person}{Abhishek Das},
  \bibinfo{person}{Ramakrishna Vedantam}, \bibinfo{person}{Devi Parikh}, {and}
  \bibinfo{person}{Dhruv Batra}.} \bibinfo{year}{2020}\natexlab{}.
\newblock \showarticletitle{Grad-CAM: Visual Explanations from Deep Networks
  via Gradient-Based Localization.}
\newblock \bibinfo{journal}{\emph{Int. J. Comput. Vis.}}
  (\bibinfo{year}{2020}).
\newblock


\bibitem[Tian et~al\mbox{.}(2021)]%
        {Tian2021}
\bibfield{author}{\bibinfo{person}{Yu Tian}, \bibinfo{person}{Yuhao Yang},
  \bibinfo{person}{Xudong Ren}, \bibinfo{person}{Pengfei Wang},
  \bibinfo{person}{Fangzhao Wu}, \bibinfo{person}{Qian Wang}, {and}
  \bibinfo{person}{Chenliang Li}.} \bibinfo{year}{2021}\natexlab{}.
\newblock \showarticletitle{{Joint Knowledge Pruning and Recurrent Graph
  Convolution for News Recommendation}}.
\newblock \bibinfo{journal}{\emph{SIGIR 2021 - Proceedings of the 44th
  International ACM SIGIR Conference on Research and Development in Information
  Retrieval}} (\bibinfo{year}{2021}), \bibinfo{pages}{51--60}.
\newblock
\showISBNx{9781450380379}
\urldef\tempurl%
\url{https://doi.org/10.1145/3404835.3462912}
\showDOI{\tempurl}


\bibitem[Vaswani et~al\mbox{.}(2017)]%
        {Vaswani_Shazeer_Parmar_Uszkoreit_Jones_Gomez_Kaiser_Polosukhin_2017}
\bibfield{author}{\bibinfo{person}{Ashish Vaswani}, \bibinfo{person}{Noam
  Shazeer}, \bibinfo{person}{Niki Parmar}, \bibinfo{person}{Jakob Uszkoreit},
  \bibinfo{person}{Llion Jones}, \bibinfo{person}{Aidan~N. Gomez},
  \bibinfo{person}{Lukasz Kaiser}, {and} \bibinfo{person}{Illia Polosukhin}.}
  \bibinfo{year}{2017}\natexlab{}.
\newblock \showarticletitle{Attention is All you Need.}. In
  \bibinfo{booktitle}{\emph{Advances in Neural Information Processing Systems
  30: Annual Conference on Neural Information Processing Systems 2017, December
  4-9, 2017, Long Beach, CA, USA}}.
\newblock


\bibitem[Verma et~al\mbox{.}(2020)]%
        {Verma_Gulati_Goel_Shah_2020}
\bibfield{author}{\bibinfo{person}{Dhruv Verma}, \bibinfo{person}{Kshitij
  Gulati}, \bibinfo{person}{Vasu Goel}, {and} \bibinfo{person}{Rajiv~Ratn
  Shah}.} \bibinfo{year}{2020}\natexlab{}.
\newblock \showarticletitle{Fashionist: Personalising Outfit Recommendation for
  Cold-Start Scenarios.}. In \bibinfo{booktitle}{\emph{MM ’20: The 28th ACM
  International Conference on Multimedia, Virtual Event / Seattle, WA, USA,
  October 12-16, 2020}}.
\newblock


\bibitem[Wang et~al\mbox{.}(2020b)]%
        {Wang2020}
\bibfield{author}{\bibinfo{person}{Chong Wang}, \bibinfo{person}{Lisa Kim},
  \bibinfo{person}{Grace Bang}, \bibinfo{person}{Himani Singh},
  \bibinfo{person}{Russell Kociuba}, \bibinfo{person}{Steven Pomerville}, {and}
  \bibinfo{person}{Xiaomo Liu}.} \bibinfo{year}{2020}\natexlab{b}.
\newblock \showarticletitle{{Discovery news: A generic framework for financial
  news recommendation}}.
\newblock \bibinfo{journal}{\emph{AAAI 2020 - 34th AAAI Conference on
  Artificial Intelligence}} (\bibinfo{year}{2020}),
  \bibinfo{pages}{13390--13395}.
\newblock
\showISBNx{9781577358350}
\showISSN{2159-5399}
\urldef\tempurl%
\url{https://doi.org/10.1609/aaai.v34i08.7054}
\showDOI{\tempurl}


\bibitem[Wang et~al\mbox{.}(2020a)]%
        {Wang2020_Next}
\bibfield{author}{\bibinfo{person}{Jianling Wang}, \bibinfo{person}{Kaize
  Ding}, \bibinfo{person}{Liangjie Hong}, \bibinfo{person}{Huan Liu}, {and}
  \bibinfo{person}{James Caverlee}.} \bibinfo{year}{2020}\natexlab{a}.
\newblock \showarticletitle{{Next-item Recommendation with Sequential
  Hypergraphs}}.
\newblock \bibinfo{journal}{\emph{SIGIR 2020 - Proceedings of the 43rd
  International ACM SIGIR Conference on Research and Development in Information
  Retrieval}} (\bibinfo{year}{2020}), \bibinfo{pages}{1101--1110}.
\newblock
\showISBNx{9781450380164}
\urldef\tempurl%
\url{https://doi.org/10.1145/3397271.3401133}
\showDOI{\tempurl}


\bibitem[Wang et~al\mbox{.}(2019b)]%
        {Wang_Jiang_Xu_Xie_2019}
\bibfield{author}{\bibinfo{person}{Peng Wang}, \bibinfo{person}{Yunsheng
  Jiang}, \bibinfo{person}{Chunxu Xu}, {and} \bibinfo{person}{Xiaohui Xie}.}
  \bibinfo{year}{2019}\natexlab{b}.
\newblock \showarticletitle{Overview of Content-Based Click-Through Rate
  Prediction Challenge for Video Recommendation.}. In
  \bibinfo{booktitle}{\emph{Proceedings of the 27th ACM International
  Conference on Multimedia, MM 2019, Nice, France, October 21-25, 2019}}.
\newblock


\bibitem[Wang et~al\mbox{.}(2021a)]%
        {Wang2021}
\bibfield{author}{\bibinfo{person}{Wenjie Wang}, \bibinfo{person}{Fuli Feng},
  \bibinfo{person}{Xiangnan He}, \bibinfo{person}{Liqiang Nie}, {and}
  \bibinfo{person}{Tat~Seng Chua}.} \bibinfo{year}{2021}\natexlab{a}.
\newblock \showarticletitle{{Denoising Implicit Feedback for Recommendation}}.
\newblock \bibinfo{journal}{\emph{WSDM 2021 - Proceedings of the 14th ACM
  International Conference on Web Search and Data Mining}}
  (\bibinfo{year}{2021}), \bibinfo{pages}{373--381}.
\newblock
\showISBNx{9781450382977}
\showISSN{2331-8422}
\urldef\tempurl%
\url{https://doi.org/10.1145/3437963.3441800}
\showDOI{\tempurl}
\showeprint[arxiv]{2006.04153}


\bibitem[Wang et~al\mbox{.}(2019a)]%
        {Wang_He_Wang_Feng_Chua_2019}
\bibfield{author}{\bibinfo{person}{Xiang Wang}, \bibinfo{person}{Xiangnan He},
  \bibinfo{person}{Meng Wang}, \bibinfo{person}{Fuli Feng}, {and}
  \bibinfo{person}{Tat-Seng Chua}.} \bibinfo{year}{2019}\natexlab{a}.
\newblock \showarticletitle{Neural Graph Collaborative Filtering.}. In
  \bibinfo{booktitle}{\emph{Proceedings of the 42nd International ACM SIGIR
  Conference on Research and Development in Information Retrieval, SIGIR 2019,
  Paris, France, July 21-25, 2019.}}
\newblock


\bibitem[Wang et~al\mbox{.}(2021b)]%
        {Wang2021_Count}
\bibfield{author}{\bibinfo{person}{Zhenlei Wang}, \bibinfo{person}{Jingsen
  Zhang}, \bibinfo{person}{Hongteng Xu}, \bibinfo{person}{Xu Chen},
  \bibinfo{person}{Yongfeng Zhang}, \bibinfo{person}{Wayne~Xin Zhao}, {and}
  \bibinfo{person}{Ji~Rong Wen}.} \bibinfo{year}{2021}\natexlab{b}.
\newblock \showarticletitle{{Counterfactual Data-Augmented Sequential
  Recommendation}}.
\newblock \bibinfo{journal}{\emph{SIGIR 2021 - Proceedings of the 44th
  International ACM SIGIR Conference on Research and Development in Information
  Retrieval}} (\bibinfo{year}{2021}), \bibinfo{pages}{347--356}.
\newblock
\showISBNx{9781450380379}
\urldef\tempurl%
\url{https://doi.org/10.1145/3404835.3462855}
\showDOI{\tempurl}


\bibitem[Wei et~al\mbox{.}(2020)]%
        {Wei_Wang_Nie_He_Chua_2020}
\bibfield{author}{\bibinfo{person}{Yinwei Wei}, \bibinfo{person}{Xiang Wang},
  \bibinfo{person}{Liqiang Nie}, \bibinfo{person}{Xiangnan He}, {and}
  \bibinfo{person}{Tat-Seng Chua}.} \bibinfo{year}{2020}\natexlab{}.
\newblock \showarticletitle{Graph-Refined Convolutional Network for Multimedia
  Recommendation with Implicit Feedback.}. In \bibinfo{booktitle}{\emph{MM
  ’20: The 28th ACM International Conference on Multimedia, Virtual Event /
  Seattle, WA, USA, October 12-16, 2020}}.
\newblock


\bibitem[Wei et~al\mbox{.}(2019)]%
        {Wei_Wang_Nie_He_Hong_Chua_2019}
\bibfield{author}{\bibinfo{person}{Yinwei Wei}, \bibinfo{person}{Xiang Wang},
  \bibinfo{person}{Liqiang Nie}, \bibinfo{person}{Xiangnan He},
  \bibinfo{person}{Richang Hong}, {and} \bibinfo{person}{Tat-Seng Chua}.}
  \bibinfo{year}{2019}\natexlab{}.
\newblock \showarticletitle{MMGCN: Multi-modal Graph Convolution Network for
  Personalized Recommendation of Micro-video.}. In
  \bibinfo{booktitle}{\emph{Proceedings of the 27th ACM International
  Conference on Multimedia, MM 2019, Nice, France, October 21-25, 2019}}.
\newblock


\bibitem[Wu et~al\mbox{.}(2019)]%
        {Wu_Wu_An_Huang_Huang_Xie_2019}
\bibfield{author}{\bibinfo{person}{Chuhan Wu}, \bibinfo{person}{Fangzhao Wu},
  \bibinfo{person}{Mingxiao An}, \bibinfo{person}{Jianqiang Huang},
  \bibinfo{person}{Yongfeng Huang}, {and} \bibinfo{person}{Xing Xie}.}
  \bibinfo{year}{2019}\natexlab{}.
\newblock \showarticletitle{NPA: Neural News Recommendation with Personalized
  Attention.}. In \bibinfo{booktitle}{\emph{Proceedings of the 25th ACM SIGKDD
  International Conference on Knowledge Discovery \& Data Mining, KDD 2019,
  Anchorage, AK, USA, August 4-8, 2019.}}
\newblock


\bibitem[Wu et~al\mbox{.}(2020)]%
        {Wu_Qiao_Chen_Wu_Qi_Lian_Liu_Xie_Gao_Wu_et_2020}
\bibfield{author}{\bibinfo{person}{Fangzhao Wu}, \bibinfo{person}{Ying Qiao},
  \bibinfo{person}{Jiun-Hung Chen}, \bibinfo{person}{Chuhan Wu},
  \bibinfo{person}{Tao Qi}, \bibinfo{person}{Jianxun Lian},
  \bibinfo{person}{Danyang Liu}, \bibinfo{person}{Xing Xie},
  \bibinfo{person}{Jianfeng Gao}, \bibinfo{person}{Winnie Wu}, {and}
  \bibinfo{person}{et al.}} \bibinfo{year}{2020}\natexlab{}.
\newblock \showarticletitle{MIND: A Large-scale Dataset for News
  Recommendation.}. In \bibinfo{booktitle}{\emph{Proceedings of the 58th Annual
  Meeting of the Association for Computational Linguistics, ACL 2020, Online,
  July 5-10, 2020}}.
\newblock


\bibitem[Xie et~al\mbox{.}(2020)]%
        {Xie_Sun_Liu_Gao_Ding_Cui_2020}
\bibfield{author}{\bibinfo{person}{Xu Xie}, \bibinfo{person}{Fei Sun},
  \bibinfo{person}{Zhaoyang Liu}, \bibinfo{person}{Jinyang Gao},
  \bibinfo{person}{Bolin Ding}, {and} \bibinfo{person}{Bin Cui}.}
  \bibinfo{year}{2020}\natexlab{}.
\newblock \showarticletitle{Contrastive Pre-training for Sequential
  Recommendation.}
\newblock \bibinfo{journal}{\emph{CoRR}} (\bibinfo{year}{2020}).
\newblock


\bibitem[Yang et~al\mbox{.}(2020)]%
        {Yang_Xie_Wang_Yuan_Ding_Yan_2020}
\bibfield{author}{\bibinfo{person}{Xuewen Yang}, \bibinfo{person}{Dongliang
  Xie}, \bibinfo{person}{Xin Wang}, \bibinfo{person}{Jiangbo Yuan},
  \bibinfo{person}{Wanying Ding}, {and} \bibinfo{person}{Pengyun Yan}.}
  \bibinfo{year}{2020}\natexlab{}.
\newblock \showarticletitle{Learning Tuple Compatibility for Conditional Outfit
  Recommendation.}. In \bibinfo{booktitle}{\emph{MM ’20: The 28th ACM
  International Conference on Multimedia, Virtual Event / Seattle, WA, USA,
  October 12-16, 2020}}.
\newblock


\bibitem[Yu et~al\mbox{.}(2020)]%
        {Yu_Gan_Wei_Cheng_Nie_2020}
\bibfield{author}{\bibinfo{person}{Xuzheng Yu}, \bibinfo{person}{Tian Gan},
  \bibinfo{person}{Yinwei Wei}, \bibinfo{person}{Zhiyong Cheng}, {and}
  \bibinfo{person}{Liqiang Nie}.} \bibinfo{year}{2020}\natexlab{}.
\newblock \showarticletitle{Personalized Item Recommendation for Second-hand
  Trading Platform.}. In \bibinfo{booktitle}{\emph{MM ’20: The 28th ACM
  International Conference on Multimedia, Virtual Event / Seattle, WA, USA,
  October 12-16, 2020}}.
\newblock


\bibitem[Zhang et~al\mbox{.}(2021a)]%
        {ZhangJD0YCTHWHC21}
\bibfield{author}{\bibinfo{person}{Ningyu Zhang}, \bibinfo{person}{Qianghuai
  Jia}, \bibinfo{person}{Shumin Deng}, \bibinfo{person}{Xiang Chen},
  \bibinfo{person}{Hongbin Ye}, \bibinfo{person}{Hui Chen},
  \bibinfo{person}{Huaixiao Tou}, \bibinfo{person}{Gang Huang},
  \bibinfo{person}{Zhao Wang}, \bibinfo{person}{Nengwei Hua}, {and}
  \bibinfo{person}{Huajun Chen}.} \bibinfo{year}{2021}\natexlab{a}.
\newblock \showarticletitle{AliCG: Fine-grained and Evolvable Conceptual Graph
  Construction for Semantic Search at Alibaba}. In
  \bibinfo{booktitle}{\emph{The 27th ACM SIGKDD Conference on Knowledge
  Discovery and Data Mining, Virtual Event, Singapore, August 14-18, 2021}}.
\newblock


\bibitem[Zhang et~al\mbox{.}(2018)]%
        {Zhang_Wu_Zhu_2018}
\bibfield{author}{\bibinfo{person}{Quanshi Zhang}, \bibinfo{person}{Ying~Nian
  Wu}, {and} \bibinfo{person}{Song-Chun Zhu}.} \bibinfo{year}{2018}\natexlab{}.
\newblock \showarticletitle{Interpretable Convolutional Neural Networks.}. In
  \bibinfo{booktitle}{\emph{2018 IEEE Conference on Computer Vision and Pattern
  Recognition, CVPR 2018, Salt Lake City, UT, USA, June 18-22, 2018}}.
\newblock


\bibitem[Zhang et~al\mbox{.}(2021b)]%
        {Zhang2021}
\bibfield{author}{\bibinfo{person}{Shengyu Zhang}, \bibinfo{person}{Dong Yao},
  \bibinfo{person}{Zhou Zhao}, \bibinfo{person}{Tat~Seng Chua}, {and}
  \bibinfo{person}{Fei Wu}.} \bibinfo{year}{2021}\natexlab{b}.
\newblock \showarticletitle{{CauseRec: Counterfactual User Sequence Synthesis
  for Sequential Recommendation}}.
\newblock \bibinfo{journal}{\emph{SIGIR 2021 - Proceedings of the 44th
  International ACM SIGIR Conference on Research and Development in Information
  Retrieval}} (\bibinfo{year}{2021}), \bibinfo{pages}{367--377}.
\newblock
\showISBNx{9781450380379}
\urldef\tempurl%
\url{https://doi.org/10.1145/3404835.3462908}
\showDOI{\tempurl}
\showeprint[arxiv]{2109.05261}


\bibitem[Zhang et~al\mbox{.}(2019)]%
        {Zhang_Yuan_Li_Zhang_2019}
\bibfield{author}{\bibinfo{person}{Xinran Zhang}, \bibinfo{person}{Xin Yuan},
  \bibinfo{person}{Yunwei Li}, {and} \bibinfo{person}{Yanru Zhang}.}
  \bibinfo{year}{2019}\natexlab{}.
\newblock \showarticletitle{Cold-Start Representation Learning: A
  Recommendation Approach with Bert4Movie and Movie2Vec.}. In
  \bibinfo{booktitle}{\emph{Proceedings of the 27th ACM International
  Conference on Multimedia, MM 2019, Nice, France, October 21-25, 2019}}.
\newblock


\bibitem[Zhou et~al\mbox{.}(2020)]%
        {zhou2020contrastive}
\bibfield{author}{\bibinfo{person}{Chang Zhou}, \bibinfo{person}{Jianxin Ma},
  \bibinfo{person}{Jianwei Zhang}, \bibinfo{person}{Jingren Zhou}, {and}
  \bibinfo{person}{Hongxia Yang}.} \bibinfo{year}{2020}\natexlab{}.
\newblock \showarticletitle{Contrastive Learning for Debiased Candidate
  Generation in Large-Scale Recommender Systems}.
\newblock \bibinfo{journal}{\emph{arXiv preprint cs.IR/2005.12964}}
  (\bibinfo{year}{2020}).
\newblock


\end{thebibliography}

\end{document}